\documentclass[aps,pra,superscriptaddress,twocolumn]{revtex4-2}

\usepackage{graphicx}
\usepackage{amsmath}
\usepackage{amssymb}
\usepackage{hyperref}
\usepackage[utf8]{inputenc}
\usepackage{mathtools}
\usepackage[english]{babel}
\usepackage{bbm}
\hypersetup{colorlinks=true, linkcolor=blue, citecolor=blue, urlcolor=blue}
\usepackage{xcolor}
\usepackage{braket}
\usepackage[normalem]{ulem}

\newcommand{\ie}{i.\,e.,\ }

\newcommand{\fref}[1]{\text{Fig.}~\ref{#1}}

\newcommand{\eref}[1]{\text{Eq.}~\eqref{#1}}

\begin{document}
\title{Characterizing superradiant dynamics in atomic arrays via a cumulant expansion approach}

\author{Oriol Rubies-Bigorda}
\email{orubies@mit.edu}
\thanks{equal contributor}
\affiliation{Physics Department, Massachusetts Institute of Technology, Cambridge, Massachusetts 02139, USA}
\affiliation{Department of Physics, Harvard University, Cambridge, Massachusetts 02138, USA}
\author{Stefan Ostermann}
\email{stefanostermann@g.harvard.edu}
\thanks{equal contributor}
\affiliation{Department of Physics, Harvard University, Cambridge, Massachusetts 02138, USA}
\author{Susanne F. Yelin}
\affiliation{Department of Physics, Harvard University, Cambridge, Massachusetts 02138, USA}

\begin{abstract}
Ordered atomic arrays with subwavelength lattice spacing emit light collectively. For fully inverted atomic arrays, this results in an initial burst of radiation and a fast build up of coherences between the atoms at initial times. Based on a cumulant expansion of the equations of motion, we derive exact analytical expressions for the emission properties and numerically analyze the full many-body problem resulting in the collective decay process for unprecedented system sizes of up to a few hundred atoms. We benchmark the cumulant expansion approach and show that it correctly captures the cooperative dynamics resulting in superradiance. For fully inverted arrays, this allows us to extract the scaling of the superradiant peak with particle number. For partially excited arrays where no coherences are shared among atoms, we also determine the critical number of excitations required for the emergence of superradiance in one- and two-dimensional geometries. In addition, we study the robustness of superradiance in the case of non-unit filling and position disorder.
\end{abstract}

\maketitle

\section{Introduction}

\begin{figure}
    \centering
    \includegraphics[width = \columnwidth]{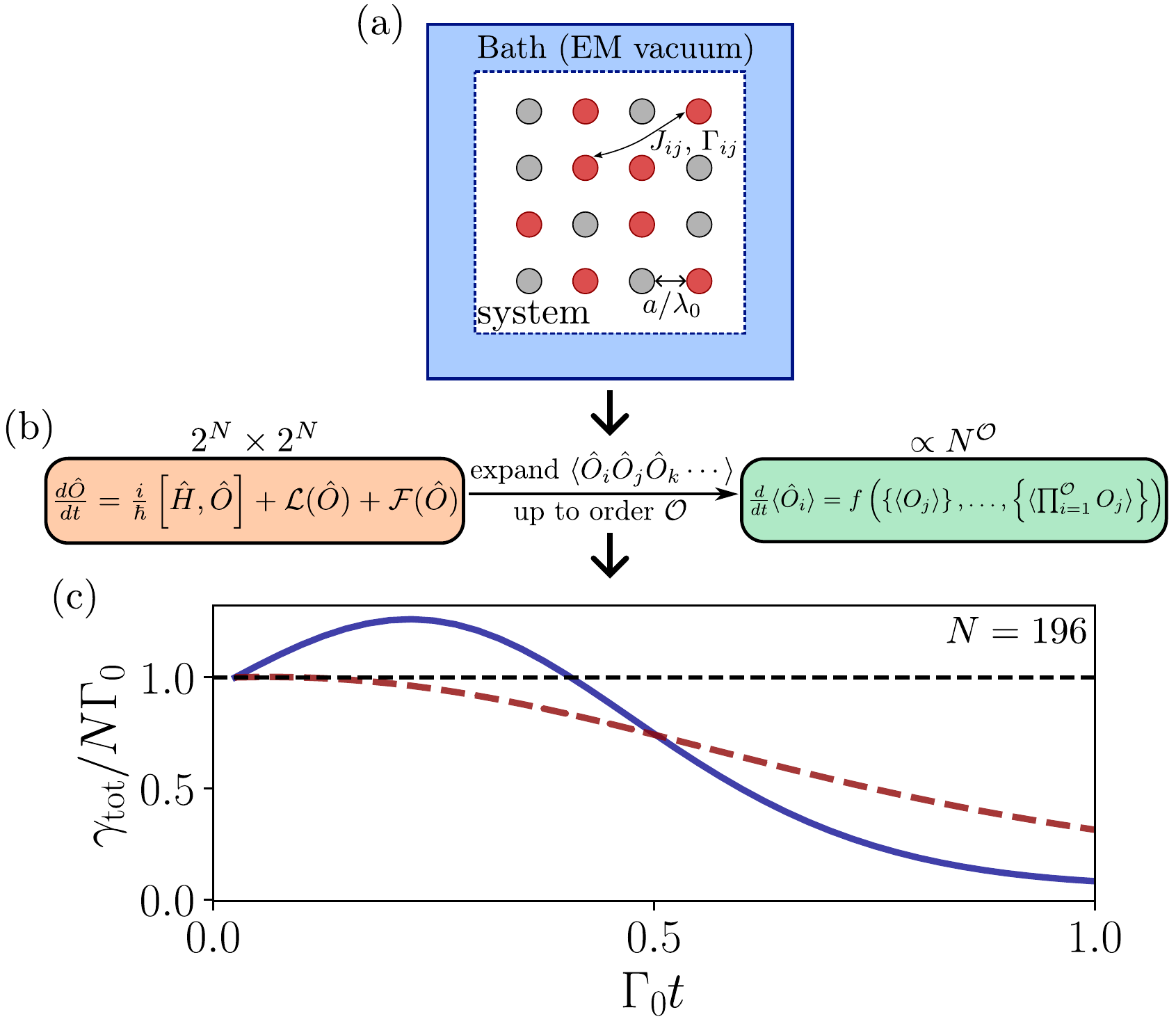}
    \caption{Schematic illustration of the theoretical approach. (a) The EM vacuum mediates interactions among emitters trapped in a periodic array. Superradianct dynamics is characterized for fully or partially excited arrays where no coherences are shared among emitters at $t=0$. (b) The complexity of the system can be reduced by truncating the number of equations of motion via a cumulant expansion of multi-order correlators. (c) This approach allows the characterization of superradiant dynamics for large system sizes. The bottom panel shows the total emission rate $\gamma_\mathrm{tot}/(N\Gamma_0)$ of an examplary time evolution for a chain of $N=196$ atoms and lattice spacings $a=0.3\lambda_0$ (solid line) and $a=0.5\lambda_0$ (dashed line).}
    \label{fig:schematic}
\end{figure}

The interaction of dense atomic ensembles with light gives rise to a plethora of interesting many-body effects. One paradigmatic example is Dicke superradiance~\cite{Dicke_originalpaper,gross_haroche}, a phenomenon in which the atoms in a totally inverted point-like sample synchronize and emit light coherently. This results in the cooperative speed-up of the atomic decay process and the emergence of a \textit{superradiant} radiation burst [see Fig.~\ref{fig:schematic}(c)]. Various aspects of Dicke superradiance have been studied \cite{Super_theory_Rehler,super_dicke_bonifacio,super_dicke_buhmann,super_dicke_liberato,super_dicke_malz,super_dicke_molmer} and experimentally observed for a broad variety of platforms, ranging from atomic gases~\cite{exp_super_atoms_1,exp_super_atoms_2,exp_super_atoms_3,exp_super_atoms_4,exp_super_atoms_5,exp_super_atoms_6,exp_super_atoms_7} to solid state systems such as quantum dots~\cite{experiment_quantumdots_1,experiment_quantumdots_2}, nitrogen-vacancy centres~\cite{experiment_NV} and two-dimensional materials~\cite{experiment_2dmat}.

Over the last years, ordered atomic arrays with subwavelength lattice spacing have emerged as a promising platform to study collective light-matter coupling in free space. In the few excitation limit, these systems naturally exhibit collective superradiant and subradiant states~\cite{AnaAsenjo_2017_general}. Various aspects of this low-excitation regime were studied in detail in recent years~\cite{AnaAsenjo_2017_general,Bloch_mirror,Ephi_2017,PRL_CAdams,Oriol_subradiance,Ruostekoski_subradiance_PRX,Ritsch_subradiance_9fold,Janos_2017_PRL,PRA_motion_Ephi,MorenoCardoner_Rydberg,Samuel_PRA,Zhang2022photonphoton}. However, analyzing the cooperative dynamics for large system sizes in the multi-excitation regime is challenging due to the rapidly growing Hilbert space. This restricts the study of ordered atomic ensembles with multiple excitations to very small systems of about ten atoms, for which the full solution of the open-system master equation is still feasible by means of Monte Carlo wave function (MCWF) methods~\cite{few_atoms_charmichael,few_atoms_charmichael_2,Ana_superradiance_1}.

For fully inverted atomic ensembles, it has been recently shown that the existence of a superradiant burst can be simply determined from the the statistics of the first two emitted photons~\cite{Ana_superradiance_1,Ana_superradiance_2,Ana_superradiance_3, Robicheaux_superradiance}. This method does not require to propagate the equations of motion in time, and therefore allows the efficient analysis of very large systems (up to $10^6$ atoms). It provides, however, no information about the decay dynamics, such as the magnitude of the superradiant peak or the existence of subradiance at late times. Alternatively, recent theoretical studies have developed an effective two-atom description of the many-body system capable of capturing superradiance in inverted three- and two-dimensional ensembles~\cite{AMO_GuindarLin, Oriol_superradiance,Hanzhen}.

In this work, we employ cumulant expansions of the operator expectation values governing the system's dynamics to characterize the decay process of dipole-dipole coupled arrays of atoms~\cite{Cumulant_Kubo,Ritsch_cumulants_dipole,Ritsch_cumulants_package, Robicheaux_cumulants}. Based on neglecting high order quantum correlations, cumulant or cluster expansions drastically reduce the degrees of freedom used to describe the system. As opposed to MCWF methods~\cite{dum_monte_1992, molmer_monte_1993}, which require $2^N$ variables to describe an $N$-atom array, the cumulant expansion up to order $n$ contains only $\sim N^n$ terms (see sketch in~\fref{fig:schematic}). To this end, we derive the equations of motion for second and third order cumulant expansions in Section~\ref{section: model}, which provide analytical insights on the mechanisms leading to superradiance. We then benchmark the formalism in Section~\ref{section: benchmarking} by comparing the resulting dynamics with that predicted by MCWF. We demonstrate that cumulant expansions capture the early dynamics resulting in the superradiant burst with remarkable accuracy and find that they allow to simulate unprecedented system sizes of up to a few hundred atoms, more than one order of magnitude larger than with MCWF.

Based on these insights we then employ this formalism to unveil new insights on the physics of superradiant emission (section~\ref{section: superradiance}). In particular, we characterize the magnitude of the superradiant peak and analyze its scaling as a function of lattice spacing and atom number for large one-dimensional and two-dimensional inverted arrays in Section~\ref{subsection: superradiance_inverted}. We additionally demonstrate that the enhanced atomic decay in inverted systems originates from a fast buildup of coherences among atoms.
Hence, a natural question arises: what is the critical fraction of excited emitters required for this cooperative synchronization effect to occur if the atoms don't share coherences at initial times? In Section~\ref{subsection: superradiance_partially_inverted}, we analytically determine this critical condition for incoherently partially inverted arrays, and further compute the value of the superradiant peak versus excitation fraction using cumulant expansions.
Finally, we extend the study to atomic arrays with finite filling fractions in Section~\ref{subsection: superradiance_missing_atoms} and with position disorder in Section~\ref{subsection: superradiance_position_disorder}.

\section{Model}
\label{section: model}
We consider an ensemble of $N$ identical two-level atoms that interacts with the vacuum electromagnetic field in free space. Applying the Markov approximation and integrating out the field degrees of freedom (see~\fref{fig:schematic}), one obtains the equations of motion for the atomic operators in the Heisenberg picture \cite{Lehmberg_1970_1,Lehmberg_1970_2}
\begin{equation}
\label{eq: Heisenberg_EOM}
    \frac{d\hat{O}}{dt} = \frac{i}{\hbar} \left[ \hat{H}, \hat{O}\right] + \mathcal{L}(\hat{O}) + \mathcal{F}(\hat{O}).
\end{equation}
Here, $\hat{H}$ and $\mathcal{L}(\hat{O})$ respectively describe the coherent and incoherent dipole-dipole coupling mediated by the vacuum field 
\begin{align}
    \hat{H} &= \sum_{i,j \neq n} J_{ij}  \hat{\sigma}_i^{eg} \hat{\sigma}_j^{ge}, \\
    \mathcal{L}(\hat{O}) &= \sum_{i,j} \frac{\Gamma_{ij}}{2} \left( 2 \hat{\sigma}_i^{eg} \hat{O} \hat{\sigma}_j^{ge} - \hat{\sigma}_i^{eg} \hat{\sigma}_j^{ge} \hat{O} - \hat{O} \hat{\sigma}_i^{eg} \hat{\sigma}_j^{ge} \right),
\end{align}
where $\hat{\sigma}_i^{eg} = |e_i \rangle \langle g_i |$ and $\hat{\sigma}_i^{ge} = |g_i \rangle \langle e_i |$ are the raising and lowering operators for atom $i$, which is located at $\mathbf{r}_i$. Additionally, the coherent $J_{ij}$ and dissipative $\Gamma_{ij}$ couplings between atoms $i$ and $j$ are obtained from the Green's tensor for a point dipole in vacuum $\textbf{G}$, given in Appendix~\ref{app: Greensfunction}, as
\begin{equation}
\label{eq: shift_decay_greens}
    J_{ij} - i \Gamma_{ij}/2 = -\frac{3\pi \gamma_0}{\omega_0} {\mathbf{d}^\dagger} \textbf{G}(\textbf{r}_{ij}, \omega_0) \mathbf{d},
\end{equation}
where $\omega_0 = 2 \pi c / \lambda_0$ corresponds to the transition wavelength of the emitters, $\mathbf{d}$ to their transition dipole moment and $\mathbf{r}_{ij}= \mathbf{r}_i - \mathbf{r}_j$ to the vector connecting atoms $i$ and $j$. Here, $\Gamma_{ij}=\Gamma_0$ is given by the spontaneous decay rate of a single atom in vacuum and the Lamb-shift $J_{ii}$ is included in the definition of $\omega_0$. For the remainder of this work, we consider a transition dipole moment perpendicular to the array. Note that similar results are obtained for other polarizations.

Finally, the last term in Eq.~(\ref{eq: Heisenberg_EOM}), $\mathcal{F}(\hat{O})$, describes the quantum Langevin noise arising from vacuum fluctuations \cite{Lehmberg_1970_1}. Assuming white noise for $\mathcal{F}(\hat{O})$, the expectation value $\langle \mathcal{F}(\hat{O}) \rangle$ vanishes. Since we are ultimately  interested in averages over atomic operators $\langle \hat{O} \rangle$, we drop this term in the following discussion to simplify notation.

In this work, we consider initial states where $N_\mathrm{exc}$ atoms are incoherently excited at $t=0$,
\begin{equation}
\label{eq: incoherent_initial_condition}
    |\psi_\mathrm{incoh} \rangle = \prod_{i \in E} \hat{\sigma}_i^{eg} |g \rangle,
\end{equation}
where $|g \rangle$ is the state with all atoms in the ground state and $E$ denotes the set of initially excited atoms. The two particle correlations $\langle \hat{\sigma}_i^{eg} \hat{\sigma}_j^{ge}\rangle$ vanish for these states. Hence, no coherences are shared among atoms at initial times even for partially excited arrays. Such states can be prepared by either using a spatially incoherent light source to excite the atomic atomic array or by imposing a random detuning pattern over the duration of the excitation pulse.

The equations of motion of the relevant first- and second-order  operators $\hat{\sigma}_i^{ee}$, $\hat{\sigma}_i^{eg} \hat{\sigma}_j^{ge}$ and $\hat{\sigma}_i^{ee} \hat{\sigma}_j^{ee}$ can be readily obtained from~\eref{eq: Heisenberg_EOM}
\pagebreak
\begin{widetext}
\begin{subequations}
\label{eq: EOM_operators_upto_2ndorder}
\begin{align}
    \frac{d}{dt}\hat{\sigma}_i^{ee} &= - \Gamma_{0} \hat{\sigma}_i^{ee} + \sum_{n \neq i}  \bigg\{ \left( i J_{ni}-\frac{\Gamma_{ni}}{2} \right) \hat{\sigma}_n^{eg} \hat{\sigma}_i^{ge} + \left(-i J_{in}-\frac{\Gamma_{in}}{2} \right) \hat{\sigma}_i^{eg} \hat{\sigma}_n^{ge} \bigg\}, \\
    \frac{d}{dt} \hat{\sigma}_i^{eg} \hat{\sigma}_j^{ge} &= - \Gamma_{0} \hat{\sigma}_i^{eg} \hat{\sigma}_j^{ge} + \frac{\Gamma_{ji}}{2} (4 \hat{\sigma}_i^{ee} \hat{\sigma}_j^{ee} - \hat{\sigma}_i^{ee} - \hat{\sigma}_j^{ee} ) + i J_{ji} (\hat{\sigma}_j^{ee} - \hat{\sigma}_i^{ee}) \nonumber \\
    &+ \sum_{n \neq i,j} \bigg\{ \left(i J_{jn}+\frac{\Gamma_{jn}}{2} \right) ( 2 \hat{\sigma}_j^{ee} \hat{\sigma}_i^{eg} \hat{\sigma}_n^{ge} - \hat{\sigma}_i^{eg} \hat{\sigma}_n^{ge}) + \left(-i J_{ni}+\frac{\Gamma_{ni}}{2} \right) ( 2 \hat{\sigma}_i^{ee} \hat{\sigma}_n^{eg} \hat{\sigma}_j^{ge} - \hat{\sigma}_n^{eg} \hat{\sigma}_j^{ge}) \bigg\}, \\
    \frac{d}{dt} \hat{\sigma}_i^{ee} \hat{\sigma}_j^{ee} &= - 2 \Gamma_{0} \hat{\sigma}_i^{ee} \hat{\sigma}_j^{ee} + \sum_{n \neq i,j} \bigg\{ \left(i J_{nj}-\frac{\Gamma_{nj}}{2} \right) \hat{\sigma}_i^{ee} \hat{\sigma}_n^{eg} \hat{\sigma}_j^{ge} + \left(-i J_{jn}-\frac{\Gamma_{jn}}{2} \right) \hat{\sigma}_i^{ee} \hat{\sigma}_j^{eg} \hat{\sigma}_n^{ge} \nonumber \\
     &+ \left(i J_{ni}-\frac{\Gamma_{ni}}{2} \right) \hat{\sigma}_j^{ee} \hat{\sigma}_n^{eg} \hat{\sigma}_i^{ge} + \left(-i J_{in}-\frac{\Gamma_{in}}{2} \right) \hat{\sigma}_j^{ee} \hat{\sigma}_i^{eg} \hat{\sigma}_n^{ge} \bigg\}.
\end{align} 
\label{eq: motion_full}
\end{subequations}
\end{widetext}
Operators of order $n$,~\ie involving $n$ products of individual atomic operators, are generally coupled to operators of order $n+1$. This leads to a coupled set of differential equations for system operators up to order $N$, where $N$ denotes the particle number.

\subsection{Cumulant expansion}

The number of equations needed to fully describe the system grows exponentially with atom number, limiting the system sizes that can be numerically simulated to about sixteen atoms. To study larger systems, one can take expectation values of the equations of motion in Eq.~(\ref{eq: EOM_operators_upto_2ndorder}) and truncate the set of equations by approximating averages of higher-order operators with combinations of averages of lower-order ones (see~\fref{fig:schematic}). This method is known as cumulant or cluster expansion~\cite{Cumulant_Kubo, Ritsch_cumulants_dipole, Robicheaux_cumulants,Ritsch_cumulants_package}. For initial states given by~\eref{eq: incoherent_initial_condition}, only the three second-order operators given in~\eref{eq: motion_full} develop non-zero expectation values $\langle \hat{\sigma}_i^{ee} \rangle$, $\langle \hat{\sigma}_i^{ee} \hat{\sigma}_j^{ee} \rangle$, $\langle \hat{\sigma}_i^{eg} \hat{\sigma}_j^{ge} \rangle$ during the time evolution. Approximating the three-operator averages in Eq.~(\ref{eq: EOM_operators_upto_2ndorder}) as $\langle  \hat{\sigma}_a^{ee} \hat{\sigma}_b^{eg} \hat{\sigma}_c^{ge} \rangle = \langle  \hat{\sigma}_a^{ee} \rangle \langle \hat{\sigma}_b^{eg} \hat{\sigma}_c^{ge} \rangle $, one obtains a closed set of differential equations for the first- and second-order expectation values, which correspond to the second-order cumulant expansion (see Appendix~\ref{app: second_ordr_cumulant}).

A similar procedure can be used to derive the third-order cumulant expansion of the atomic system, which additionally includes the expectation values $\langle \hat{\sigma}_i^{ee} \hat{\sigma}_j^{ee} \hat{\sigma}_k^{ee} \rangle$ and $\langle \hat{\sigma}_i^{ee} \hat{\sigma}_j^{eg} \hat{\sigma}_k^{ge} \rangle$. The explicit form of the corresponding equations of motion is presented in Appendix~\ref{app: third_ordr_cumulant}.

\subsection{Emission properties}
\begin{figure}
    \centering
    \includegraphics[width = \columnwidth]{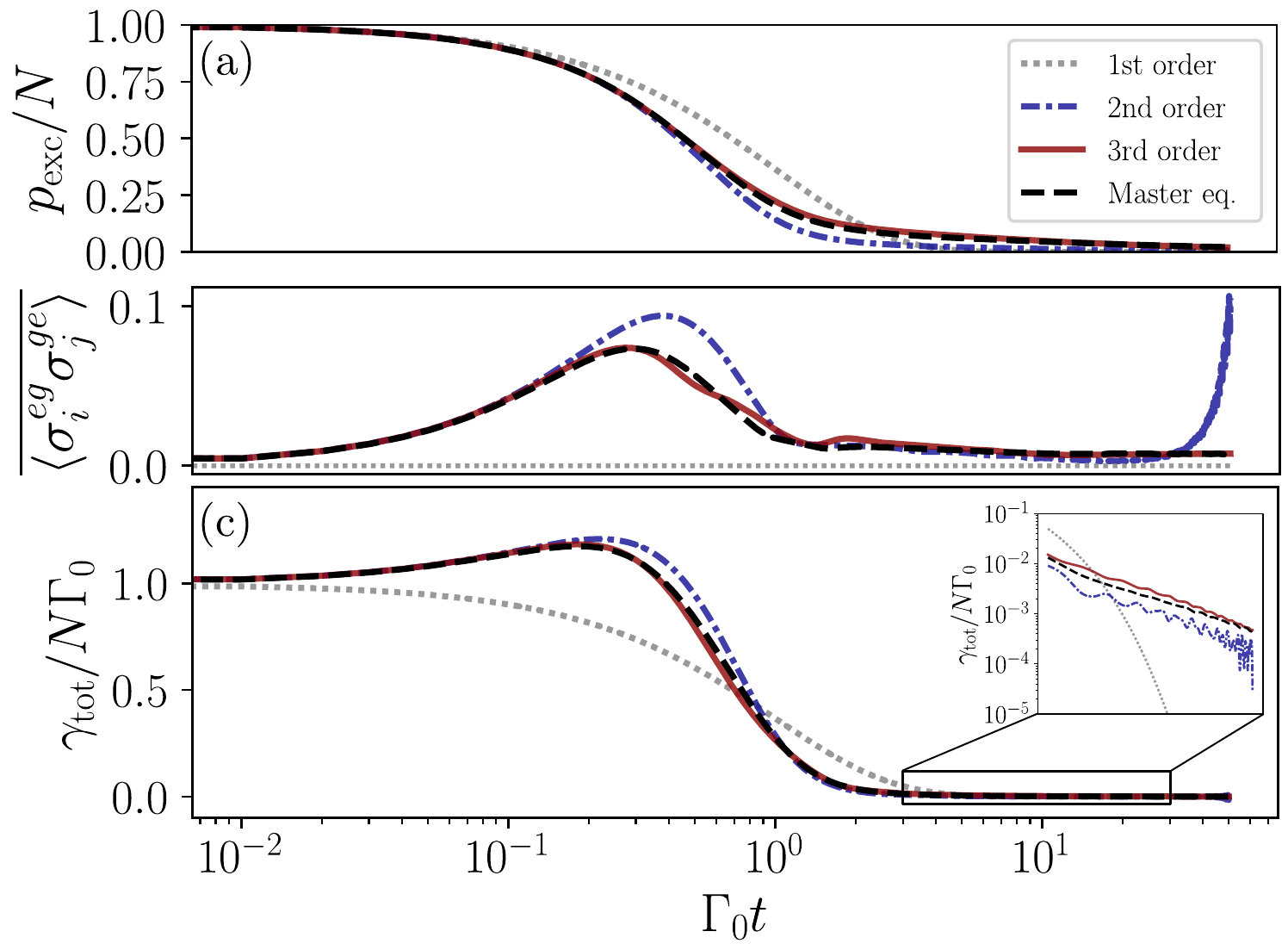}
    \caption{Comparison between the dynamics obtained via cumulant expansion of orders 1--3 and the solution of the full master equation for a chain of $N = 10$ atoms with lattice spacing $a=0.1\lambda_0$. (a) Excited state population $p_\mathrm{exc}(t)$, (b) average pair correlations and (c) total emission rate $\gamma_\mathrm{tot}/(N\Gamma_0)$. The inset in panel (c) represents the total emission rate at late times in logarithmic scale. The atoms are considered to be polarized in the direction perpendicular to the lattice plane throughout this work.}
    \label{fig: benchmark_timeevol}
\end{figure}

While a recent theoretical work~\cite{Robicheaux_cumulants} focused on employing a cumulant expansion approach to determine two-time correlation functions, we hereby focus on characterizing the emission properties of the atomic array. In fact, the equations of motion corresponding to the second order cumulant expansion provide a powerful tool to elucidate the fundamental mechanisms resulting in superradiance.

The decay process of the atomic ensemble can be characterized by the excited-state population
\begin{equation}
    p_\mathrm{exc}(t) = \sum_i \langle \hat{\sigma}_i^{ee} \rangle (t)
\end{equation}
and the total emission rate of the system
\begin{equation}
    \gamma_\mathrm{tot}(t) \equiv -\frac{d}{dt}  p_\mathrm{exc}(t) = \Gamma_{0} \sum_i \langle \hat{\sigma}_i^{ee} \rangle + \sum_{i,j \neq i} \Gamma_{ij} \langle \hat{\sigma}_i^{eg} \hat{\sigma}_j^{ge} \rangle.
    \label{eq: gamma_tot}
\end{equation}
\eref{eq: gamma_tot} provides a very nice intuitive picture on the underlying mechanisms resulting in superradiance. For independent particles ($\Gamma_{ij}=0$) the emission rate is proportional to the individual atom decay rate $\Gamma_0$. For interacting particles, however, the term $\propto \Gamma_{ij} \langle \hat{\sigma}_i^{eg} \hat{\sigma}_j^{ge}\rangle$ modifies the emission rate, and can eventually result in a superradiant burst if the interactions $\Gamma_{ij}$ and pair correlations $\langle \hat{\sigma}_i^{eg} \hat{\sigma}_j^{ge}\rangle$ are sufficiently large.

The existence of a radiation burst can be predicted from the derivative of the total emission rate at $t=0$~\cite{Robicheaux_superradiance}. For initial states given by~\eref{eq: incoherent_initial_condition}, we obtain
\begin{align}
\label{eq: condition_burst_general}
    \dot{\gamma}_{\mathrm{tot},0} &= - \Gamma_{0}^2 \sum_{i}  \langle \hat{\sigma}_i^{ee} \rangle_0 \nonumber \\
    &+ \sum_{i,j \neq i}  \Gamma_{ij} \Gamma_{ji} \left( 2 \langle \hat{\sigma}_i^{ee} \hat{\sigma}_j^{ee} \rangle_0 - \frac{\langle \hat{\sigma}_i^{ee} \rangle_0 + \langle \hat{\sigma}_j^{ee} \rangle_0}{2} \right),
\end{align}
where $\langle \hat{O} \rangle_0 \equiv \langle \hat{O} \rangle (t=0)$. For $\dot{\gamma}_{\mathrm{tot},0} > 0$, the emission rate initially increases and the dynamics consequently result in a superradiant peak. For $\dot{\gamma}_{\mathrm{tot},0} < 0$, on the other hand, $\gamma_\mathrm{tot}$ initially decreases, indicating that the dissipative interactions between emitters cannot build the amount of coherences required for the appearance of a burst. For an initially fully inverted system with $\langle \hat{\sigma}_i^{ee} \rangle_0 = \langle \hat{\sigma}_i^{ee} \hat{\sigma}_j^{ee} \rangle_0 =1$, Eq.~(\ref{eq: condition_burst_general}) reduces to 
\begin{equation}
\label{eq: condition_burst_fully_inverted}
    \dot{\gamma}_{\mathrm{tot},0} = - N \Gamma_{0}^2 + \sum_{i,j \neq i}  \Gamma_{ij} \Gamma_{ji}
\end{equation}
and is identical to the expressions derived in Ref.~\cite{Robicheaux_superradiance}, as well as to the expression obtained via the two-photon correlation function in Ref.~\cite{Ana_superradiance_1}. Note that both Eqs.~(\ref{eq: condition_burst_general}) and (\ref{eq: condition_burst_fully_inverted}) are exact, as second-order cumulants perfectly capture two-photon processes at initial times.

While typically sufficient when studying the decay of fully inverted systems or partially excited coherent spin-states~\cite{Ana_superradiance_1}, this criterion can fail to identify radiation bursts for more general initial conditions. One example are incoherently driven, partially excited arrays, whose initial states are in a superposition of dark and bright states as it is presented in Ref.~\cite{our_PRL_subradiance}.
In this case, strong coherent dipole-dipole interactions $J_{ij}$ can generate a net population transfer from subradiant to superradiant states, which ultimately leads to This phenomenon, which is triggered by the Hamiltonian evolution of the system, only shows signatures in the emission rate for $t>0$. It can consequently not be captured by~\eref{eq: condition_burst_general}, which solely relies on the dissipative couplings and the atomic properties at zero time.

\section{Benchmarking cumulant expansions}
\label{section: benchmarking}
When simulating the dynamics of coupled spin systems, there is always a trade-off between the maximum number of particles that can be numerically simulated and the accuracy of the resulting dynamics. Increasing the order of the cumulant expansion, for example, reduces the former and generally improves the latter. In this section, we study which expansion order is required to correctly capture the evolution of the atomic ensemble. This analysis can only be performed for small system sizes, for which a full solution of the master equation is possible.

In~\fref{fig: benchmark_timeevol}, we plot dynamic properties of a fully inverted ten-atom chain with lattice spacing $a=0.1 \lambda_0$ using first (grey dotted line), second (blue dash-dotted line) and third (red solid line) order cumulant expansions, as well as the solution of the full master equation (black dashed). Third order cumulants exhibit very good agreement with the exact time evolution and properly capture the magnitude of the superradiant peak (maximum of $\gamma_\mathrm{tot}/N$), while second order cumulants slightly overestimate it. The outburst of radiation is accompanied by a buildup of the atomic coherences $\langle \hat{\sigma}_i^{eg} \hat{\sigma}_j^{ge} \rangle$ at early times due to the repeated application of the same bright jump operators~\cite{Ana_superradiance_1}.
\begin{figure}
    \centering
    \includegraphics[width =\columnwidth]{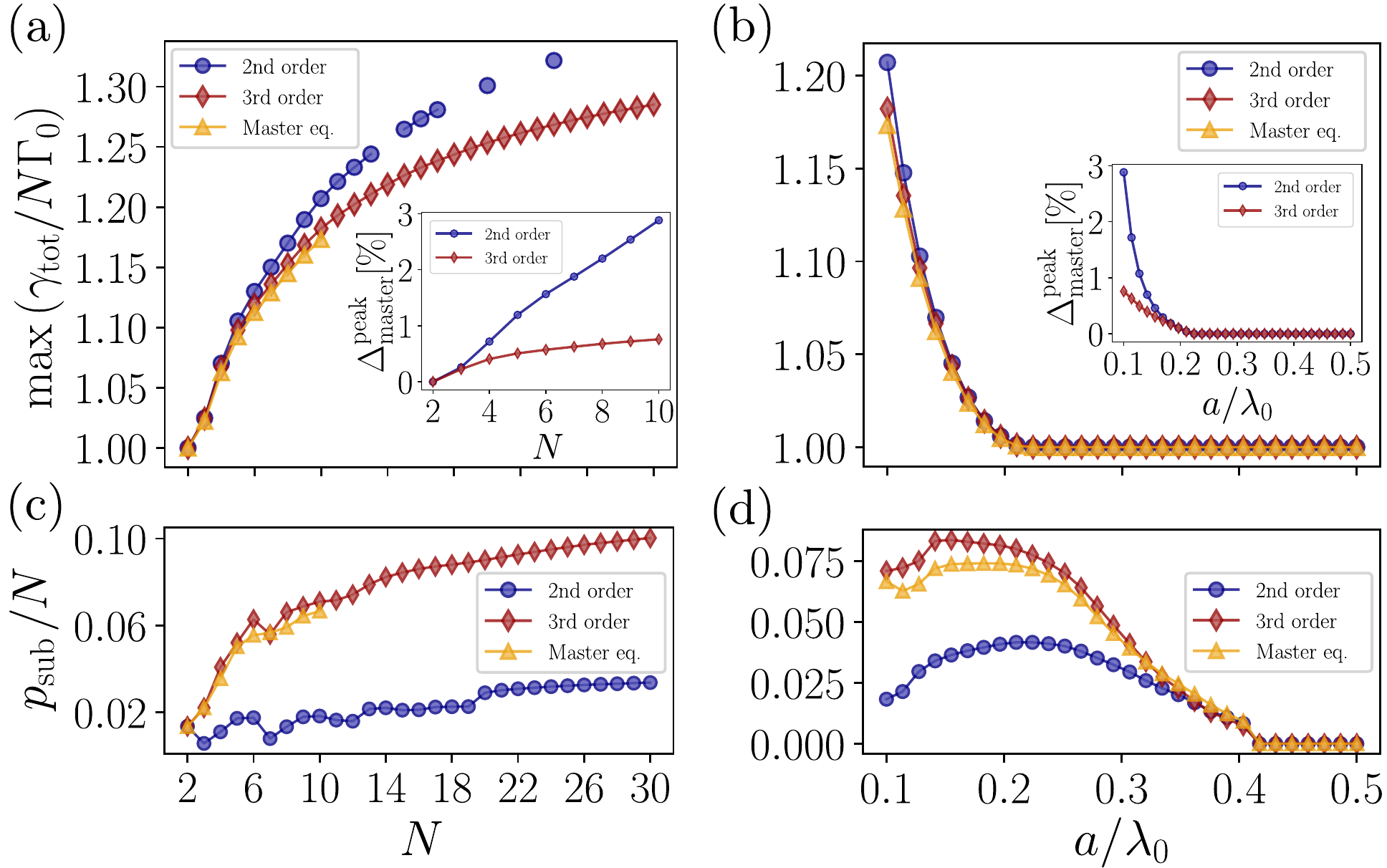}
    \caption{Benchmarking the cumulant expansion method. (a) Maximum value of the superradiant peak for a fully inverted chain of atoms with lattice spacing $a=0.1\lambda_0$ as a function of atom number. (b) Value of the superradiant peak for a $N=10$ atom chain as a function of lattice spacing.  The insets in panels (a) and (b) show the relative error between the exact master equation solution and the cumulant expansions in percent. (c)-(d) Magnitude of the subradiant population $p_\mathrm{sub}$, \ie the excited population left in the array by the time the instantaneous decay rate $\gamma_\mathrm{inst} = \gamma_\mathrm{tot}/(p_\mathrm{exc} \Gamma_0) = 0.1$, as a function of (c) particle number and (d) lattice spacing. Same parameters as in panels (a) and (b).}
    \label{fig:benchmark_Na_dep}
\end{figure}
At late times, however, the performance of the cumulant expansion worsens. In particular, the second order cumulant expansion fails to correclty predict the late dynamics [see inset in~\fref{fig: benchmark_timeevol}(c)], and the third order cumulant expansion sometimes results in unphysical behaviors such as growing excitation number in the absence of drive~\cite{Robicheaux_cumulants}. This poor performance at late times can be attributed to the population of subradiant states during the decay process. These states are typically highly entangled, and therefore contain many high-order inter-atom coherences that lead to destructive interference of the electromagnetic field emitted by different atoms. Consequently, an expansion that neglects higher-order coherences performs worse in predicting these quantities. For very large system sizes ($N>100$) and small lattice spacings, the cumulant expansion can also result in unphysical behaviors such as a growing excited state population in the absence of drive.

Finally, it is worth noting that the first-order cumulant expansion, commonly referred to as the mean-field approximation, simply results in an independent exponential decay of the atomic ensemble (without a buildup of the atomic coherences) and does not capture neither super- nor subradiance.

To benchmark the accuracy of cumulant expansions of different orders, we compare two figures of merit of the decay of inverted atomic arrays as a function of particle number $N$ and lattice spacing $a$. In Fig.~\ref{fig:benchmark_Na_dep}(a), we plot the maximum emission rate or magnitude of the superradiant peak, $\max ( \gamma_\mathrm{tot}/N )$, as a function of particle number $N$ for a one-dimensional chain with lattice spacing $a=0.1 \lambda_0$. The inset shows the error made by second-order (blue circles) and third-order (red diamonds) cumulant expansions, defined as $\Delta_\mathrm{master}^\mathrm{peak} = \max ( \gamma_\mathrm{tot}^\mathrm{cum})/ \max ( \gamma_\mathrm{tot}^\mathrm{master}) - 1$, in percent. Not surprisingly, third-order cumulant expansions are more accurate than second-order ones and, more importantly, their error does \emph{not} grow linearly with $N$. Fig.~\ref{fig:benchmark_Na_dep}(b) demonstrates that the accuracy considerably worsens with decreasing lattice spacing. This effect can be attributed to the rapidly growing energy shifts, which generate correlations between atoms.

Another relevant quantity that characterizes the decay process of the ensemble is the subradiant population $p_\mathrm{sub}$, that is, the amount of excitation left in the atomic array once the system enters the subradiant regime. This occurs once the instantaneous decay rate $\gamma_\mathrm{inst}\equiv \gamma_\mathrm{tot}/(p_\mathrm{exc} \Gamma_0)$ goes below $0.1$, as defined in Ref.~\cite{our_PRL_subradiance}. Again, third-order cumulants provide an excellent estimate of this quantity, while second-order cumulants fail to capture it. The larger errors of second-order cumulants when estimating $p_\mathrm{sub}$ as opposed to $\max ( \gamma_\mathrm{tot}/N )$ are due to the fact that the former contains information of the decay process until times much longer than the radiation burst, where subradiant evolution starts to dominate. Still, $p_\mathrm{sub}$ largely depends on the superradiant dynamics and is therefore accurately captured by the third-order cumulant expansions~\cite{our_PRL_subradiance}, even if the subsequent subradiant evolution is not [see inset of~\fref{fig: benchmark_timeevol}(c)].

\section{Superradiance}
\label{section: superradiance}

Having elucidated the accuracy with which second and third order cumulant expansions capture superradiance, we now investigate the conditions under which  this phenomenon emerges in one- and two-dimensional atomic arrays, as well as the magnitude of the radiation burst reached by the ensemble of emitters.

\subsection{Fully inverted arrays}
\label{subsection: superradiance_inverted}

\begin{figure}
    \centering
    \includegraphics[width = \columnwidth]{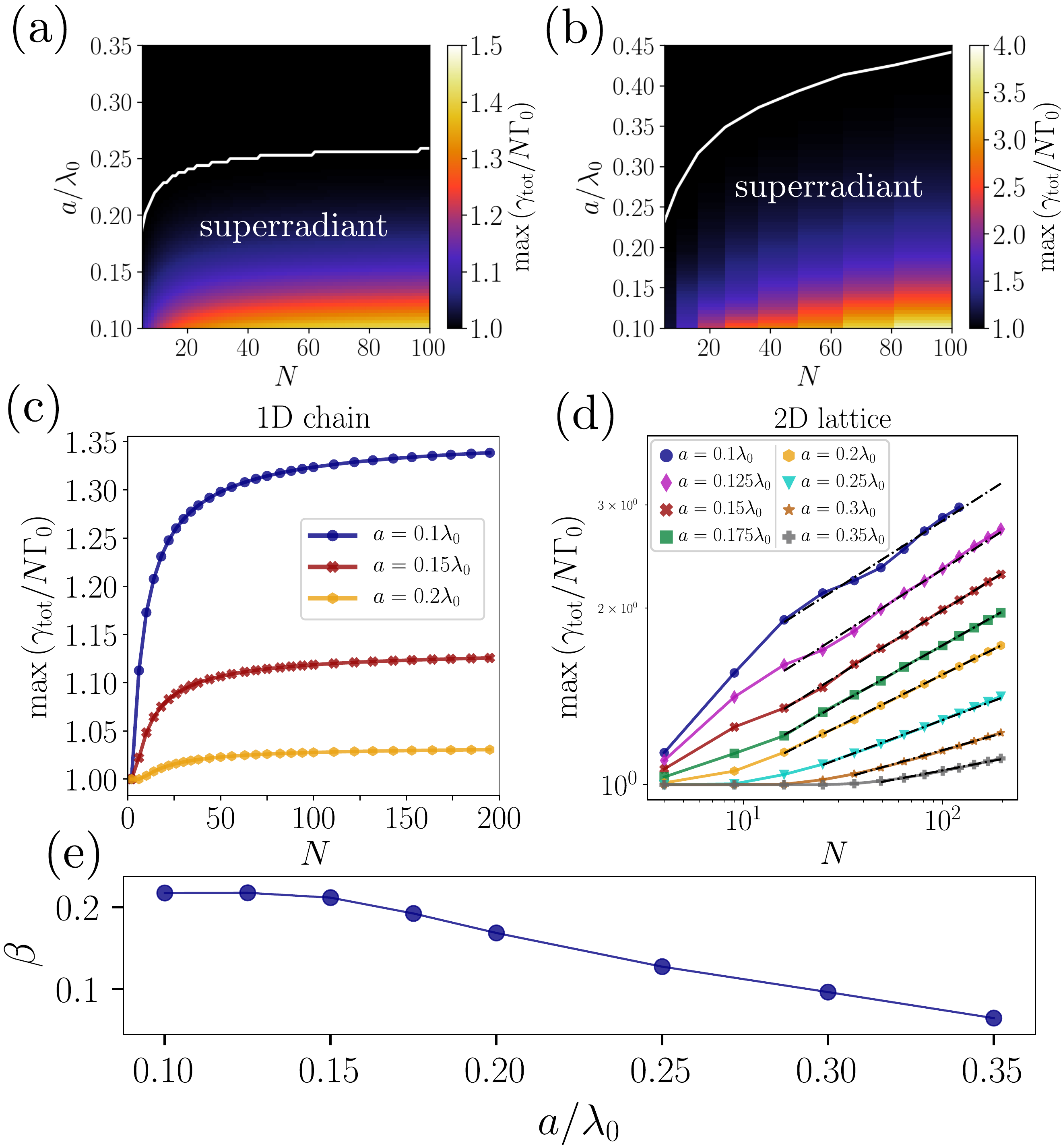}
    \caption{(a)-(b) Maximum emission rate $\max ( \gamma_\mathrm{tot}/N\Gamma_0)$ for (a) an atomic chain and (b) a two-dimensional square lattice as a function of particle number $N$ and spacing $a/\lambda_0$ obtained via a second order cumulant expansion. The white solid line separates the region where a superradiant burst is observed from the region where no superradiance occurs. (c)-(d) Scaling of the superradiant peak as a function of particle number for  (c) a chain and (d) a square lattice for different lattice spacings. (e) Exponent $\beta$ characterizing the power law dependence of the superradiant peak with particle number, $\propto N^\beta$, for a square lattice. The values are obtained from the linear fits in panel (d), shown as dash dotted lines.}
    \label{fig:phase_diag}
\end{figure}
Superradiant emission is typically studied in fully inverted arrays, where all atoms have been initially excited by an intense laser drive. In this scenario, the existence of a burst as a function of system size and lattice spacing can be inferred from Eq.~(\ref{eq: condition_burst_fully_inverted})~\cite{Ana_superradiance_1,Ana_superradiance_2,Ana_superradiance_3,Robicheaux_superradiance}. This condition is only based on the structure of the equations of motion at $t=0$ and no predictions on the resultant time evolution can be made. Here, we use the cumulant expansions approach outlined above to extend previous studies~\cite{Ana_superradiance_1, Ana_superradiance_2, Ana_superradiance_3, Hanzhen, Oriol_superradiance, Robicheaux_superradiance, Robicheaux_cumulants} on superradiance in ordered subwavelength arrays by analyzing the time evolution leading to superradiance. This allows us to quantify the magnitude of the superradiant peak for system sizes of a few hundred atoms, compatible with state of the art experiments.

In~\fref{fig:phase_diag} we determine the superradiant regime for different particle numbers and lattice spacings for a chain of atoms [\fref{fig:phase_diag}(a)] and a square lattice [\fref{fig:phase_diag}(b)]. We calculate the time evolution for a certain set of parameters ($N$ and $a$) by solving the cumulant equations up to second order and determine the maximum of the total emission rate, which is plotted as a color code in~\fref{fig:phase_diag}(a) and (b). A superradiant peak occurs if $\max(\gamma_\mathrm{tot})>N\gamma_0$, that is, if the maximum emission rate is larger than that of independent decay. The white solid line separates the regime where a superradiant burst occurs from the regime where no cooperative enhancement of the atomic emission takes place, and coincides with the critical value obtained from ~\eref{eq: condition_burst_fully_inverted}. The results presented in~\fref{fig:phase_diag}(a) and (b) demonstrate that the transition between both regimes is not sharp and suggest that lattice spacings well below the ones predicted in Ref.~\cite{Ana_superradiance_1,Ana_superradiance_2,Ana_superradiance_3,Robicheaux_superradiance} [white line~\fref{fig:phase_diag}(a) and (b)] are required to experimentally observe the initial speed-up of radiation.

\fref{fig:phase_diag}(c) and (d) show the scaling of the superradiant peak with particle number for a 1D chain and a square lattice, respectively ( obtained with a third order cumulant expansion).  We find that the maximum emission rate saturates for $N\rightarrow \infty$ for a one-dimensional chain, whereas it increases as $\propto N^\beta$ for a two-dimensional configuration.  Performing linear fits of the curves shown in~\fref{fig:phase_diag}(d), we find that the exponent $\beta$ increases for decreasing lattice spacing until it saturates for $a/\lambda_0\rightarrow 0$ [see~\fref{fig:phase_diag}(e)]. It is worth noting that the values of $\beta$ found in this work are consistent with those reported in Ref.~\cite{Oriol_superradiance}.

\subsection{Partially inverted arrays}
\label{subsection: superradiance_partially_inverted}

\begin{figure}
    \centering
    \includegraphics[width = \columnwidth]{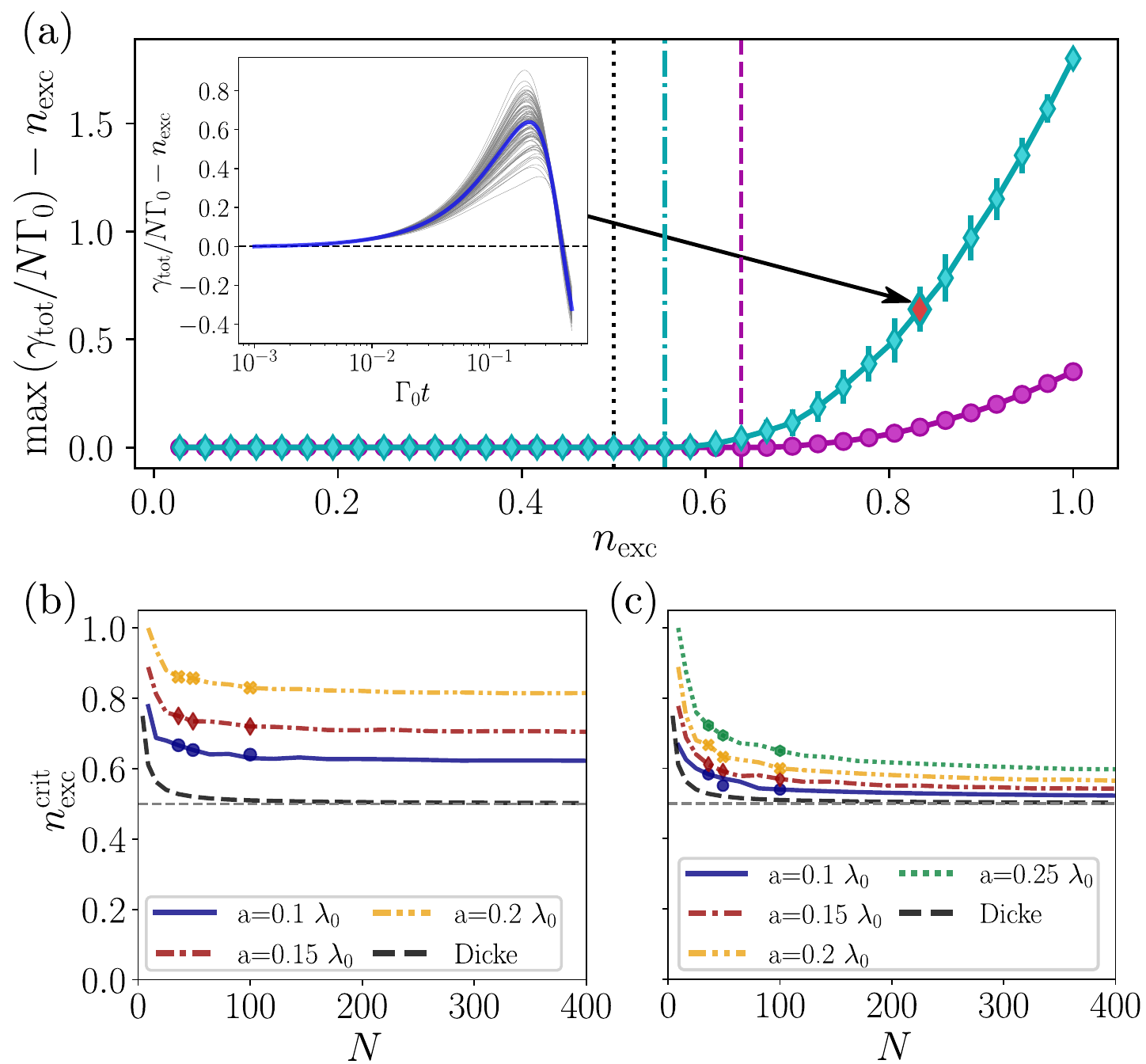}
    \caption{(a) Value of the superradiant peak as a function of the initial incoherent excitation fraction for $N=36$ atoms. Magenta diamonds correspond to a one dimensional chain and cyan diamonds to a two dimensional square lattice, both with spacing $a/\lambda_0 = 0.1$. The inset shows the time evolution of 100 trajectories, each of them corresponding to a different, random initial distribution of thirty excitations. The blue curve indicates the average over all trajectories and is used to determine the magnitude of the superradiant peak shown in the main figure. The vertical dashed and dashdotted lines correspond to the critical excitation fraction, determined by~\eref{eq: critical_excitation_deexcited} for atomic chains and two-dimensional lattices, respectively. (b)-(c) Critical excitation, obtained via~\eref{eq: critical_excitation_deexcited}, as a function of particle number for (b) one-dimensional and (c) two-dimensional arrays with different lattice spacings. For $a/\lambda_0 \rightarrow 0$ the system approaches the Dicke case (black dashed line). The blue circles ($a/\lambda_0=0.1$), red diamonds ($a/\lambda_0=0.15$), yellow crosses ($a/\lambda_0=0.2$) and green hexagons ($a/\lambda_0=0.25$) are obtained numerically and show a good agreement with the analytical result.}
    \label{fig: partial_excitation}
\end{figure}

For fully inverted arrays, the mechanism resulting in superradiance relies on the fast build up of coherences between pairs of atoms [see~\eref{eq: gamma_tot} and~\fref{fig: benchmark_timeevol}(b)]. The natural question arises: how many atoms need to be excited for these coherences to build up fast enough  and for a superradiant burst to occur? For that, we consider atomic ensembles driven by an incoherent driving field, such that only $N_\mathrm{exc}$ randomly selected emitters are excited. This type of initial state, where $\langle \hat{\sigma}_i^{eg} \hat{\sigma}_j^{ge}\rangle_0=0$,  can be generally described by Eq.~(\ref{eq: incoherent_initial_condition}).
In the inset of Fig.~\ref{fig: partial_excitation}(a), we show in grey the emission rate for different initialization of a square lattice with thirty-six atoms, thirty of which are initially excited. While the exact magnitude of the burst depends on the specific configuration of excited atoms, one can extract a characteristic peak size by averaging over all trajectories (thick blue line). The resulting average peak size is then plotted as a function of the fraction of excited atoms, $n_\mathrm{exc} \coloneqq N_\mathrm{exc}/N$, in Fig.~\ref{fig: partial_excitation}(a). We find that there is a critical excitation fraction, $n_\mathrm{exc}^\mathrm{crit}$, required to obtain a superradiant burst both for atomic chains and two-dimensional arrays, which depends both on system size and the specific geometry.

One can analytically compute $n_\mathrm{exc}^\mathrm{crit}$ by averaging Eq.~(\ref{eq: condition_burst_general}) over all possible configurations of excited atoms. Defining $N_\mathrm{de}=N-N_\mathrm{exc}$ as the number of de-excited atoms, we obtain an average derivative of the emission rate at $t=0$ equal to (for details see Appendix~\ref{app: critical_fraction})
\begin{align}
\label{eq: deriv_tot_em_deexcited}
    \bar{\dot{\gamma}}_{\mathrm{tot},0}  =  &- N_\mathrm{exc} \Gamma_0^2  \nonumber \\
    &+ \left( 1 \! - \!\frac{3 N_\mathrm{de}}{N}  \!+ \! \frac{2 N_\mathrm{de} \left(N_\mathrm{de} -1\right)}{N(N-1)}  \right) \! \! \! \sum_{n,m \neq n} \Gamma_{ij} \Gamma_{ji}.
\end{align}
A superradiant burst is observed for $\bar{\dot{\gamma}}_{\mathrm{tot},0} > 0$, that is, for
\begin{equation}
\label{eq: critical_excitation_deexcited}
    n_\mathrm{exc} >\frac{1}{2} + \frac{1}{2N}+ \frac{N-1}{2 \sum_{n,m \neq n}  \Gamma_{ij} \Gamma_{ji}/\Gamma_0^2}.
\end{equation}
This analytical result shows perfect agreement with the numerically obtained results [see vertical dashed and dash-dotted lines in~\fref{fig: partial_excitation}(a)].

In the Dicke limit $a\rightarrow 0$, where all emitters are located at the same spatial position, all dissipative couplings are equal to the spontaneous decay rate,~\ie $\Gamma_{mn} = \Gamma_0$ $\forall m,n$, and Eq.~(\ref{eq: critical_excitation_deexcited}) results in $n_\mathrm{exc}^\mathrm{crit} = 1/2 + 1/N$. That is, an excitation fraction larger than one half,~\ie $N_\mathrm{exc}>N/2 + 1$, is required to observe a burst. This can be intuitively understood by realizing that the symmetric Dicke state with maximum decay rate corresponds to the state where half of the atoms are excited. Hence, an initially fully inverted system dynamically evolves into states with higher and higher decay rates while cascading down the Dicke ladder, which ultimately results in the appearance of a superradiant emission peak. Once half of the atomis are de-excited this trend reverses and the system starts populating states with decreasing decay rates, causing the peak to vanish. If only half (or less) of the atoms are excited initially, the system will never decay into states with a larger decay rate than the initial one, and the atomic ensemble is bound to decay exponentially without the emergence of a burst.

The analytical result in~\eref{eq: critical_excitation_deexcited} provides insight into the role of the lattice geometry and dimensionality for arbitrary system sizes. For extended arrays of arbitrary dimension, the sum in the denominator of Eq.~(\ref{eq: critical_excitation_deexcited}) is always smaller than for the Dicke case, and a larger excitation fraction is needed to attain superradiance. In one-dimensional arrays, the critical excitation fraction tends to a constant for large particle number, as shown in Fig.~\ref{fig: partial_excitation}(b). In two-dimensional arrays [see Fig.~\ref{fig: partial_excitation}(c)], however, it decays logarithmically with system size $N$, $\lim_{N \rightarrow \infty} n_\mathrm{exc}^\mathrm{crit}/N \sim 1/2 + A/ \ln (\sqrt{N})$, and eventually reaches the Dicke limit for infinite systems. For completeness, it can be shown that this scaling is improved in three-dimensional lattices, where $\lim_{N \rightarrow \infty} n_\mathrm{exc}^\mathrm{crit}/N \sim 1/2+A \times N^{-1/3}$~\cite{Ana_superradiance_3,Robicheaux_superradiance}.
Figs.~\ref{fig: partial_excitation}(b) and (c) also show that a larger excitation fraction is needed to attain superradiance in atomic chains than in a square lattice for equal atom number $N$ and lattice spacing $a$, as well as the fact that smaller values of $a$ result in a smaller $n_\mathrm{exc}^\mathrm{crit}$.

Hence, Fig.~\ref{fig: partial_excitation} and Eq.~(\ref{eq: critical_excitation_deexcited}) show that there exists a well-defined critical light intensity of the excitation pulse for which superradiance emerges. It is worth noting, however, that Eq.~(\ref{eq: critical_excitation_deexcited}) can fail to identify superradiant bursts triggered by the Hamiltonian dynamics at times later than $t=0$ (for details see Refs.~\cite{our_PRL_subradiance, AMO_GuindarLin}). While these may result in superradiant peaks appearing for $n_\mathrm{exc}<0.5$, they require strong coherent couplings $J_{ij}$ or large dephasing or Doppler broadening and therefore do not typically appear for lattice spacings $a \geq 0.1\lambda_0$ considered in Fig.~\ref{fig: partial_excitation}.

Finally, it is worth noting that, in experimental implementations of superradiance, the excitation pulse might not be able to fully invert the atom system. Our results show that superradiance is reasonably robust to finite excitation fraction, as a radiation outburst still occurs at low enough lattice spacings even if only $90 \%$ or less of the atoms are initially excited.

\subsection{Finite filling fractions}
\label{subsection: superradiance_missing_atoms}
\begin{figure}
    \centering
    \includegraphics[width=0.48\textwidth]{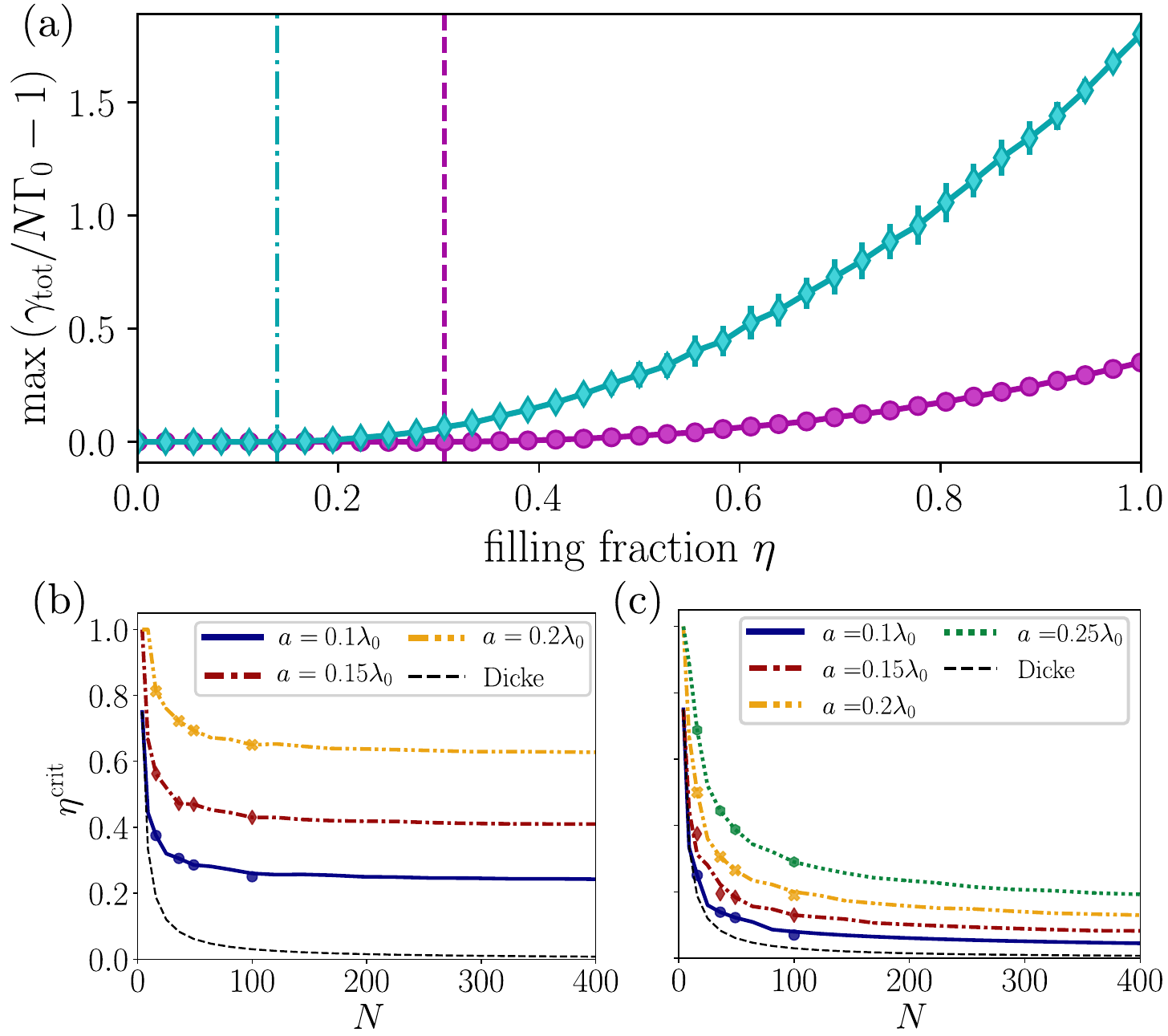}
    \caption{ (a) Value of the superradiant peak for finite filling fractions $\eta$ for a fully inverted array. Magenta diamonds correspond to a one dimensional chain and cyan diamonds to a two dimensional square lattice, both with spacing $a/\lambda_0 = 0.1$. (b)-(c) Critical filling fraction to observe a superradiant peak, obtained via~\eref{eq: excitation_number_missing_atoms}, as a function of particle number for (b) one-dimensional and (c) two-dimensional arrays with different lattice spacings. For $a/\lambda_0 \rightarrow 0$, the system approaches the Dicke case (black dashed line). The blue circles ($a/\lambda_0=0.1$), red diamonds ($a/\lambda_0=0.15$), yellow crosses ($a/\lambda_0=0.2$) and green hexagons ($a/\lambda_0=0.25$) are obtained numerically and show a good agreement with the analytical result.}
    \label{fig: missing_atoms}
\end{figure}

Typically, experimental implementations of atomic arrays also suffer from a finite filling fraction of the lattice. In Ref.~\cite{Bloch_mirror}, for example, which demonstrated that two-dimensional array with subwavelength spacing can reflect incident beams of low intensities, approximate filling fractions of ninety percent were achieved. Here, we show that superradiance for fully inverted arrays is very robust to missing atoms and that large superradiant bursts can still be achieved for filling fractions much lower than those attainable in state of the art experiments.

Fig.~\ref{fig: missing_atoms}(a) shows a similar analysis as it was performed in the previous section. Now, however, all atoms are initially excited and only a fraction $\eta\equiv N_\mathrm{filled}/N$ of the $N$ available lattice sites is populated by an excited atom. In other words, the lattice contains $N_\mathrm{hol}=N-N_\mathrm{filled}$ holes or missing atoms. \fref{fig: missing_atoms} shows the average magnitude of the peak for one- and two-dimensional arrays with $N=36$ lattice sites as a function of the filling fraction $\eta$. The peak values are again obtained by averaging over time evolutions for 100 random distributions of holes. The magnitude of the peak slowly decreases with diminishing $\eta$, and finally reaches zero at a critical filling fraction $\tilde{n}_\mathrm{exc}^\mathrm{crit}$ that depends on the properties of the atomic ensemble. 

Again, we can estimate $\eta^\mathrm{crit}$
by averaging $\dot{\gamma}_0$ over all possible configurations of missing atoms (for details see Appendix~\ref{app: critical_filling_fraction})
\begin{align}
\label{eq: deriv_tot_em_holes}
    \bar{\dot{\gamma}}_{\mathrm{tot},0}  =  &-N_\mathrm{filled} \Gamma_0^2\nonumber \\
    &+ \left( 1  -  \frac{2 N_\mathrm{hol}}{N}  +  \frac{ N_\mathrm{hol} (N_\mathrm{hol} -1)}{N(N-1)} \right)  \sum_{i,j \neq i}  \Gamma_{ij} \Gamma_{ji}.
\end{align}
On average, a burst occurs for $\bar{\dot{\gamma}}_{\mathrm{tot},0}>0$, that is, for
\begin{equation}
\label{eq: excitation_number_missing_atoms}
    \eta > \frac{1}{N} + \frac{N-1}{\sum_{i,j \neq i}  \Gamma_{ij} \Gamma_{ij}/\Gamma_0^2} .
\end{equation}
In the Dicke limit, Eq.~(\ref{eq: excitation_number_missing_atoms}) reduces to the well known expression $\eta > 2/N$, \ie $N_\mathrm{filled}>2$. For two atoms, the decay rates from $|ee\rangle$ to $(|eg\rangle + |ge\rangle)/\sqrt{2}$ and from $(|eg\rangle + |ge\rangle)/\sqrt{2}$ to $|gg\rangle$ are both $2 \Gamma_0$. The two-atom system therefore decays faster than in vacuum, but cannot develop a superradiant burst. For this to occur, the system needs to dynamically access states with larger and larger decay rates while cascading down the Dicke ladder, which naturally occurs for ensembles with three or more atoms. Additionally, it is straightforward to check that $n_\mathrm{exc}^\mathrm{crit} = 1/2 + \eta^\mathrm{crit}/2$. Thus, the scalings presented in Section~\ref{subsection: superradiance_partially_inverted} also hold for the case of missing atoms.

It is worth noting that the excitation fraction needed to observe a peak is considerably larger in the case of partially excited configurations, presented in Section~\ref{subsection: superradiance_partially_inverted}, than in the case of missing atoms or holes. The main difference between both scenarios consists on the fact that the latter remains a fully inverted system, while the former does not. That is, an inverted array with missing atoms can simply cascade down the whole ladder of superradiant states by repeated action of the brightest jump operator ---which is now weaker than in a perfect array--- and can therefore attain a burst even for very small filling fractions. However, this is not the case for incoherent partially inverted arrays, whose initial state is already a superposition of bright and dark states in an intermediate sector of the state space (see also Ref.~\cite{our_PRL_subradiance}) and therefore has a reduced likelihood to quickly build up the necessary correlations to create a superradiant burst.

\subsection{Position disorder}
\label{subsection: superradiance_position_disorder}
\begin{figure}
    \centering
    \includegraphics[width=\columnwidth]{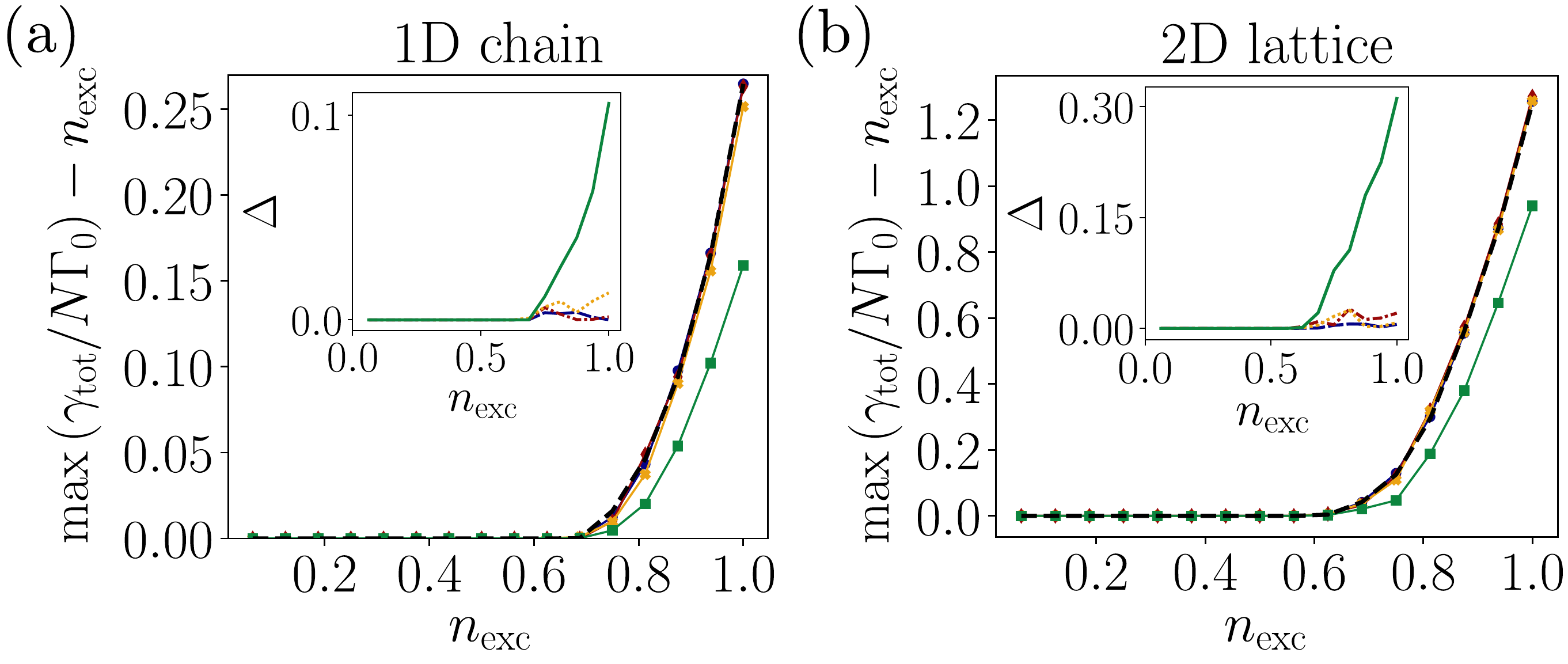}
    \caption{Effect of position disorder, sampled from a Gaussian distribution with standard deviation $\sigma = (0.01,0.05,0.1,0.2)a$ (blue circles, red diamonds, orange crosses, green squares) for a chain (a) and a square lattice (b) of $N=16$ atoms and averaged over $50$ different lattice configurations and initial excitation distributions. The dashed lines show the result for a perfect lattice without position disorder. The insets exhibit the difference $\Delta = \max \{\gamma_\mathrm{tot}/(N\Gamma_0) \}|_{\sigma = 0} -\max \{\gamma_\mathrm{tot}/(N\Gamma_0) \}|_{\sigma}$ between the zero disorder case and the disordered systems. Disorder affects superradiance if $\sigma>0.1a$.}
    \label{fig:pos_disorder}
\end{figure}

Another potential imperfection that might influence the superradiant burst in realistic setups is disorder in the atomic positions. This can either be due to some distortion in the underlying trapping potential or due to atomic motion in the respective trapping potentials.

To take position disorder into account, we sample the modified atomic positions from a normal distribution $\mathcal{N}_d(\mathbf{r}_0,\sigma)$,  where $\sigma$ denotes its standard deviation and $\mathbf{r}_0$ the positions of the disorder-less array. We then average over 100 different excitation and disorder patterns and plot the determined peak height in~\fref{fig:pos_disorder}. We find that the effect of disorder remains small as long as $\sigma$ is below $0.1a/\lambda_0$, but rapidly increases for larger $\sigma$ (see green curves in Fig.~\ref{fig:pos_disorder}). Still, these results suggest that superradiance is very robust against position disorder. 

\section{Conclusions}

We performed a numerical and analytical in-depth analysis of superradiance in ordered atomic arrays based on a cumulant expansion approach. This formalism allowed us to simulate the dynamics of open quantum systems for large particle numbers and to gain insights on the physics behind super- and subradiance. In particular, we identify the scaling of the superradiant peak with particle number both for one-dimensional and two-dimensional arrays. Additionally, we show that there exists a critical excitation fraction above which a superradiant burst occurs, and demonstrate that superradiance is a robust phenomenon that prevails in the presence of position disorder and finite filling and excitation fraction.

The results presented here are expected  to be readily observable in various state-of-the-art platforms, ranging from atomic tweezers~\cite{tweezers_Columbia} and optical lattices of cold atoms~\cite{Bloch_mirror} to solid state platforms such as two-dimensional materials~\cite{experiment_2dmat,trond_andersen} or vacancy centers in crystals~\cite{experiment_NV}. Our work also paves the way towards efficiently calculating dynamic light emission patterns from atomic arrays, as well as towards devising novel schemes to prepare multi-excitation subradiant states~\cite{our_PRL_subradiance}. Finally, the existence of a critical excitation fraction to attain superradiance exhibits a certain resemblance with the superradiant transition in the driven dissipative Dicke model~\cite{dicke_cavity_proposal_Carmichael,baumann2010dicke,dynamical_phase_transitions_dicke_Hemmerich,ferioli2022observation}. Exploring this connection is an exiting avenue for future work.

\begin{acknowledgments}
We would like to thank Valentin Walther and Yidan Wang for fruitful discussions.
O.R.B. acknowledges support from Fundación Mauricio y Carlota Botton and from Fundació Bancaria “la Caixa” (LCF/BQ/AA18/11680093). S.O. is supported by a postdoctoral fellowship of the Max Planck Harvard Research Center for Quantum Optics. SFY would like to acknowledge funding from NSF through the CUA PFC and the QSense QLCI as well as from AFOSR. 

O.R.B. and S.O. contributed equally to this work.
\end{acknowledgments}

\vspace{0.5cm}
\appendix
\section{Green's function}
\label{app: Greensfunction}
The Green's function for a point dipole in free space used in Eq.~(\ref{eq: shift_decay_greens}) can be written in Cartesian coordinates as \cite{GreensFunction_Chew,GreensFunction_novotny_hecht_2006}
\begin{align}
G_{\alpha \beta}(\mathbf{r},\omega) &= \frac{e^{i k r}}{4\pi r} \left[ \left( 1 + \frac{i}{kr} - \frac{1}{(kr)^2} \right) \delta_{\alpha\beta} \right. \nonumber \\
  &+ \left. \left(-1 - \frac{3i}{kr} + \frac{3}{(kr)^2} \right) \frac{r_\alpha r_\beta}{r^2} \right] + \frac{\delta_{\alpha \beta} \delta^{(3)}(\mathbf{r})}{3k^2},
\end{align}
where $k=\omega/c$, $r=|\mathbf{r}|$, and $\alpha,\beta=x,y,z$.

\section{Second-order cumulant expansion}
\label{app: second_ordr_cumulant}
For initial states with no correlations as defined in Eq.~(\ref{eq: incoherent_initial_condition}), the only nonzero expectation values of first- and second-order operators at $t=0$ are
\begin{subequations}
\label{eq: initial_conditions_2nd_order_cumulants}
\begin{align}
    \langle \hat{\sigma}_i^{ee} \rangle (t=0) &= \left \{
    \begin{aligned}
        &1, && \text{if atom}\ i \  \text{is excited}\ \\
        &0, && \text{otherwise}
    \end{aligned} \right. \\
     \langle \hat{\sigma}_i^{ee} \hat{\sigma}_j^{ee} \rangle (t=0) &= \left \{
    \begin{aligned}
        &1, && \text{if atom}\ i \ \text{and} \ j\  \text{are excited}\ \\
        &0, && \text{otherwise.}
    \end{aligned} \right. 
\end{align} 
\end{subequations}
Using Eq.~(\ref{eq: Heisenberg_EOM}), one can show that the only additional expectation values that become non-zero during the evolution of the system are $\langle \hat{\sigma}_i^{eg} \hat{\sigma}_j^{ge} \rangle$. This confirms that Eq.~(\ref{eq: EOM_operators_upto_2ndorder}) is sufficient to describe the dynamics of the atomic ensemble up to second order.

Taking expectation values in Eq.~(\ref{eq: EOM_operators_upto_2ndorder}) and replacing averages over third-order operators by \cite{Cumulant_Kubo, Ritsch_cumulants_package}
\begin{align}
\langle \hat{O}_1 \hat{O}_2 \hat{O}_3 \rangle &= \langle \hat{O}_1 \rangle \langle \hat{O}_2 \hat{O}_3 \rangle + \langle \hat{O}_2 \rangle \langle \hat{O}_1 \hat{O}_3 \rangle \nonumber \\
&+ \langle \hat{O}_3 \rangle \langle \hat{O}_1 \hat{O}_2 \rangle -2 \langle \hat{O}_1 \rangle \langle \hat{O}_2 \rangle \langle \hat{O}_3 \rangle,
\end{align}
one obtains a closed set of differential equations for the second order cumulants

\onecolumngrid
\begin{widetext}
\small
\begin{subequations}
\begin{align}
    \frac{d}{dt}\langle \hat{\sigma}_i^{ee} \rangle &= - \Gamma_{0} \langle \hat{\sigma}_i^{ee} \rangle + \sum_{n \neq i} \bigg\{  \left( i J_{ni}-\frac{\Gamma_{ni}}{2} \right) \langle \hat{\sigma}_n^{eg} \hat{\sigma}_i^{ge} \rangle + \left(-i J_{in}-\frac{\Gamma_{in}}{2} \right) \langle \hat{\sigma}_i^{eg} \hat{\sigma}_n^{ge} \rangle \bigg\}, \\
    \frac{d}{dt} \langle  \hat{\sigma}_i^{eg} \hat{\sigma}_j^{ge} \rangle  &= - \Gamma_{0} \langle  \hat{\sigma}_i^{eg} \hat{\sigma}_j^{ge} \rangle  + \frac{\Gamma_{ji}}{2} \left(4 \langle  \hat{\sigma}_i^{ee} \hat{\sigma}_j^{ee} \rangle  - \langle \hat{\sigma}_i^{ee} \rangle  - \langle  \hat{\sigma}_j^{ee} \rangle  \right) + i J_{ji} \left( \langle \hat{\sigma}_j^{ee} \rangle - \langle \hat{\sigma}_i^{ee} \rangle \right) \nonumber \\
    &+ \sum_{n \neq i,j} \bigg\{ \left(i J_{jn}+\frac{\Gamma_{jn}}{2} \right) \langle \hat{\sigma}_i^{eg} \hat{\sigma}_n^{ge} \rangle \left( 2 \langle \hat{\sigma}_j^{ee} \rangle - 1 \right)  + \left(-i J_{ni}+\frac{\Gamma_{ni}}{2} \right) \langle \hat{\sigma}_n^{eg} \hat{\sigma}_j^{ge} \rangle \left( 2 \langle \hat{\sigma}_i^{ee} \rangle - 1 \right) \bigg\}, \\
    \frac{d}{dt} \langle  \hat{\sigma}_i^{ee} \hat{\sigma}_j^{ee} \rangle  &= - 2 \Gamma_{0} \langle  \hat{\sigma}_i^{ee} \hat{\sigma}_j^{ee} \rangle + \sum_{n \neq i,j} \bigg\{ \left(i J_{nj}-\frac{\Gamma_{nj}}{2} \right) \langle \hat{\sigma}_i^{ee} \rangle \langle \hat{\sigma}_n^{eg} \hat{\sigma}_j^{ge} \rangle + \left(-i J_{jn}-\frac{\Gamma_{jn}}{2} \right) \langle \hat{\sigma}_i^{ee} \rangle \langle \hat{\sigma}_j^{eg} \hat{\sigma}_n^{ge} \rangle \nonumber \\
    &+ \left(i J_{ni}-\frac{\Gamma_{ni}}{2} \right) \langle \hat{\sigma}_j^{ee} \rangle \langle \hat{\sigma}_n^{eg} \hat{\sigma}_i^{ge} \rangle + \left(-i J_{in}-\frac{\Gamma_{in}}{2} \right) \langle \hat{\sigma}_j^{ee} \rangle \langle \hat{\sigma}_i^{eg} \hat{\sigma}_n^{ge} \rangle \bigg\}.
\end{align}
\end{subequations}
\end{widetext}
\normalsize

\section{Third-order cumulant expansion}
\label{app: third_ordr_cumulant}
The third-order cumulant expansion is obtained by deriving the Heisenberg equations of motion for operators up to third order and replacing the averages of fourth-order operators by lower-order ones via the rule \cite{Cumulant_Kubo, Ritsch_cumulants_package}
\begin{align}
\label{eq: cumulant_rule_4th}
\langle \hat{O}_1 \hat{O}_2 \hat{O}_3 \hat{O}_4 \rangle &= \langle \hat{O}_1 \rangle \langle \hat{O}_2 \hat{O}_3 \hat{O}_4 \rangle + \langle \hat{O}_2 \rangle \langle \hat{O}_1 \hat{O}_3 \hat{O}_4 \rangle \nonumber \\
&+ \langle \hat{O}_3 \rangle \langle \hat{O}_1 \hat{O}_2 \hat{O}_4 \rangle + \langle \hat{O}_4 \rangle \langle \hat{O}_1 \hat{O}_2 \hat{O}_3 \rangle \nonumber \\
&+ \langle \hat{O}_1 \hat{O}_2 \rangle \langle \hat{O}_3 \hat{O}_4 \rangle + \langle \hat{O}_1 \hat{O}_3 \rangle \langle \hat{O}_2 \hat{O}_4 \rangle \nonumber \\
&+ \langle \hat{O}_1 \hat{O}_4 \rangle \langle \hat{O}_2 \hat{O}_3 \rangle  \nonumber \\
&-2 \langle \hat{O}_1 \rangle \langle \hat{O}_2 \rangle \langle \hat{O}_3 \hat{O}_4 \rangle -2 \langle \hat{O}_1 \rangle \langle \hat{O}_3 \rangle \langle \hat{O}_2 \hat{O}_4 \rangle \nonumber \\
&-2 \langle \hat{O}_1 \rangle \langle \hat{O}_4 \rangle \langle \hat{O}_2 \hat{O}_3 \rangle -2 \langle \hat{O}_2 \rangle \langle \hat{O}_3 \rangle \langle \hat{O}_1 \hat{O}_4 \rangle \nonumber \\
&-2 \langle \hat{O}_2 \rangle \langle \hat{O}_4 \rangle \langle \hat{O}_1 \hat{O}_3 \rangle -2 \langle \hat{O}_3 \rangle \langle \hat{O}_4 \rangle \langle \hat{O}_1 \hat{O}_2 \rangle \nonumber \\
&+ 6 \langle \hat{O}_1 \rangle \langle \hat{O}_2 \rangle \langle \hat{O}_3 \rangle \langle \hat{O}_4 \rangle.
\end{align}
For the initial states given in Eq.~(\ref{eq: incoherent_initial_condition}), the non-zero expectation values at $t=0$ are those given in Eq.(~\ref{eq: initial_conditions_2nd_order_cumulants}) plus
\begin{equation}
         \langle \hat{\sigma}_i^{ee} \hat{\sigma}_j^{ee} \hat{\sigma}_k^{ee}  \rangle (t=0) = \left \{
    \begin{aligned}
        &1, && \text{if atom}\ i, \ j \ \text{and} \ k \  \text{are excited}\ \\
        &0, && \text{otherwise.}
    \end{aligned} \right. 
\end{equation}
Again, one can show that the the decay process only couples these expectation values of populations with $\langle \hat{\sigma}_i^{eg} \hat{\sigma}_j^{ge} \rangle$ and $\langle \hat{\sigma}_i^{ee} \hat{\sigma}_j^{eg} \hat{\sigma}_k^{ge} \rangle$. Performing the expansion in Eq.~(\ref{eq: cumulant_rule_4th}), one finally obtains the third-order cumulant expansion
\begin{widetext}
\small
\begin{subequations}
\begin{align}
    \frac{d}{dt} \langle \hat{\sigma}_i^{ee} \rangle &= - \Gamma_{0} \langle \hat{\sigma}_i^{ee}\rangle + \sum_{n \neq i}  \bigg\{ \left( i J_{ni}-\frac{\Gamma_{ni}}{2} \right) \langle \hat{\sigma}_n^{eg} \hat{\sigma}_i^{ge} \rangle + \left(-i J_{in}-\frac{\Gamma_{in}}{2} \right) \langle \hat{\sigma}_i^{eg} \hat{\sigma}_n^{ge} \rangle \bigg\},  \\
    \frac{d}{dt} \langle \hat{\sigma}_i^{eg} \hat{\sigma}_j^{ge} \rangle &= - \Gamma_{0} \hat{\sigma}_i^{eg} \hat{\sigma}_j^{ge} + \frac{\Gamma_{ji}}{2} \left(4 \langle \hat{\sigma}_i^{ee} \hat{\sigma}_j^{ee} \rangle - \langle \hat{\sigma}_i^{ee} \rangle - \langle \hat{\sigma}_j^{ee} \rangle \right) + i J_{ji} \left( \langle \hat{\sigma}_j^{ee} \rangle - \langle \hat{\sigma}_i^{ee} \rangle \right) \nonumber \\
    &+ \sum_{n \neq i,j} \bigg\{ \left(i J_{jn}+\frac{\Gamma_{jn}}{2} \right) \left( 2 \langle \hat{\sigma}_j^{ee} \hat{\sigma}_i^{eg} \hat{\sigma}_n^{ge} \rangle - \langle \hat{\sigma}_i^{eg} \hat{\sigma}_n^{ge} \rangle \right) + \left(-i J_{ni}+\frac{\Gamma_{ni}}{2} \right) \left( 2 \langle \hat{\sigma}_i^{ee} \hat{\sigma}_n^{eg}  \hat{\sigma}_j^{ge} \rangle - \langle \hat{\sigma}_n^{eg} \hat{\sigma}_j^{ge} \rangle \right) \bigg\}, \\
    \frac{d}{dt} \langle \hat{\sigma}_i^{ee} \hat{\sigma}_j^{ee} \rangle &= - 2 \Gamma_{0} \langle \hat{\sigma}_i^{ee} \hat{\sigma}_j^{ee} \rangle + \sum_{n \neq i,j} \bigg\{ \left(i J_{ni}-\frac{\Gamma_{ni}}{2} \right) \langle \hat{\sigma}_j^{ee} \hat{\sigma}_n^{eg} \hat{\sigma}_i^{ge} \rangle + \left(-i J_{in}-\frac{\Gamma_{in}}{2} \right) \langle \hat{\sigma}_j^{ee} \hat{\sigma}_i^{eg} \hat{\sigma}_n^{ge} \rangle + (i \leftrightarrow j) \bigg\}, \\
     \frac{d}{dt} \langle \hat{\sigma}_i^{ee} \hat{\sigma}_j^{ee} \hat{\sigma}_k^{ee} \rangle &= -3 \Gamma_0 \langle \hat{\sigma}_i^{ee} \hat{\sigma}_j^{ee} \hat{\sigma}_k^{ee} \rangle \sum_{n \neq i,j,k} \bigg\{ \nonumber \\ 
     &+ \left(i J_{ni}-\frac{\Gamma_{ni}}{2} \right) \left( \langle  \hat{\sigma}_j^{ee} \rangle \langle \hat{\sigma}_k^{ee} \hat{\sigma}_n^{eg} \hat{\sigma}_i^{ge} \rangle +  \langle \hat{\sigma}_k^{ee} \rangle \langle \hat{\sigma}_j^{ee} \hat{\sigma}_n^{eg} \hat{\sigma}_i^{ge} \rangle + \langle \hat{\sigma}_j^{ee} \hat{\sigma}_k^{ee} \rangle \langle \hat{\sigma}_n^{eg} \hat{\sigma}_i^{ge} \rangle - 2 \langle \hat{\sigma}_j^{ee} \rangle \langle \hat{\sigma}_k^{ee} \rangle \langle \hat{\sigma}_n^{eg} \hat{\sigma}_i^{ge} \rangle \right) \nonumber \\
     &+ \left(-i J_{in}-\frac{\Gamma_{in}}{2} \right) \left( \langle  \hat{\sigma}_j^{ee} \rangle \langle \hat{\sigma}_k^{ee} \hat{\sigma}_i^{eg} \hat{\sigma}_n^{ge} \rangle +  \langle \hat{\sigma}_k^{ee} \rangle \langle \hat{\sigma}_j^{ee} \hat{\sigma}_i^{eg} \hat{\sigma}_n^{ge} \rangle + \langle \hat{\sigma}_j^{ee} \hat{\sigma}_k^{ee} \rangle \langle \hat{\sigma}_i^{eg} \hat{\sigma}_n^{ge} \rangle - 2 \langle \hat{\sigma}_j^{ee} \rangle \langle \hat{\sigma}_k^{ee} \rangle \langle \hat{\sigma}_i^{eg} \hat{\sigma}_n^{ge} \rangle \right) \nonumber \\
     &+ (i \leftrightarrow j) + (i \leftrightarrow k) \bigg\},
     \\
     \frac{d}{dt} \langle \hat{\sigma}_i^{ee} \hat{\sigma}_j^{eg} \hat{\sigma}_k^{ge} \rangle &= -2 \Gamma_0 \langle \hat{\sigma}_i^{ee} \hat{\sigma}_j^{eg} \hat{\sigma}_k^{ge} \rangle + \left(-i J_{ij}-\frac{\Gamma_{ij}}{2} \right) \langle \hat{\sigma}_j^{ee} \hat{\sigma}_i^{eg} \hat{\sigma}_k^{ge} \rangle + \left(i J_{ki}-\frac{\Gamma_{ki}}{2} \right) \langle \hat{\sigma}_k^{ee} \hat{\sigma}_j^{eg} \hat{\sigma}_i^{ge} \rangle \nonumber \\
     &+ i J_{kj} \left( \langle \hat{\sigma}_i^{ee} \hat{\sigma}_k^{ee}\rangle - \langle \hat{\sigma}_i^{ee} \hat{\sigma}_j^{ee}\rangle \right) + \frac{\Gamma_{kj}}{2} \left( 4 \langle \hat{\sigma}_i^{ee} \hat{\sigma}_j^{ee} \hat{\sigma}_k^{ee} \rangle -\langle \hat{\sigma}_i^{ee} \hat{\sigma}_j^{ee}\rangle - \langle \hat{\sigma}_i^{ee} \hat{\sigma}_k^{ee}\rangle \right) \nonumber \\
     &+ \sum_{n \neq i,j,k} \bigg\{ \left(i J_{ni}-\frac{\Gamma_{ni}}{2} \right) \left( \langle  \hat{\sigma}_n^{eg} \hat{\sigma}_i^{ge} \rangle \langle \hat{\sigma}_j^{eg} \hat{\sigma}_k^{ge} \rangle + \langle  \hat{\sigma}_n^{eg} \hat{\sigma}_k^{ge} \rangle \langle \hat{\sigma}_j^{eg} \hat{\sigma}_i^{ge} \rangle \right) \nonumber \\
     &+ \left(-i J_{in}-\frac{\Gamma_{in}}{2} \right) \left( \langle  \hat{\sigma}_i^{eg} \hat{\sigma}_n^{ge} \rangle \langle \hat{\sigma}_j^{eg} \hat{\sigma}_k^{ge} \rangle + \langle  \hat{\sigma}_i^{eg} \hat{\sigma}_k^{ge} \rangle \langle \hat{\sigma}_j^{eg} \hat{\sigma}_n^{ge} \rangle \right) \nonumber \\
     &+ \left(-i J_{nj}+\frac{\Gamma_{nj}}{2} \right) \left( 2 \langle  \hat{\sigma}_i^{ee} \rangle \langle \hat{\sigma}_j^{ee} \hat{\sigma}_n^{eg} \hat{\sigma}_k^{ge} \rangle + 2 \langle  \hat{\sigma}_j^{ee} \rangle \langle \hat{\sigma}_i^{ee} \hat{\sigma}_n^{eg} \hat{\sigma}_k^{ge} \rangle + 2 \langle   \hat{\sigma}_i^{ee} \hat{\sigma}_j^{ee} \rangle \langle \hat{\sigma}_n^{eg} \hat{\sigma}_k^{ge} \rangle  \right) \nonumber \\
     &+ \left(i J_{nj}-\frac{\Gamma_{nj}}{2} \right) \left( 4 \langle  \hat{\sigma}_i^{ee} \rangle \langle  \hat{\sigma}_j^{ee} \rangle \langle \hat{\sigma}_n^{eg} \hat{\sigma}_k^{ge} \rangle + \langle \hat{\sigma}_i^{ee} \hat{\sigma}_n^{eg} \hat{\sigma}_k^{ge} \rangle   \right) \nonumber \\
     &+ \left(i J_{kn}+\frac{\Gamma_{kn}}{2} \right) \left( 2 \langle  \hat{\sigma}_i^{ee} \rangle \langle \hat{\sigma}_k^{ee} \hat{\sigma}_j^{eg} \hat{\sigma}_n^{ge} \rangle + 2 \langle  \hat{\sigma}_k^{ee} \rangle \langle \hat{\sigma}_i^{ee} \hat{\sigma}_j^{eg} \hat{\sigma}_n^{ge} \rangle + 2 \langle   \hat{\sigma}_i^{ee} \hat{\sigma}_k^{ee} \rangle \langle \hat{\sigma}_j^{eg} \hat{\sigma}_n^{ge} \rangle  \right) \nonumber \\
     &+ \left(-i J_{kn}-\frac{\Gamma_{kn}}{2} \right) \left( 4 \langle  \hat{\sigma}_i^{ee} \rangle \langle  \hat{\sigma}_k^{ee} \rangle \langle \hat{\sigma}_j^{eg} \hat{\sigma}_n^{ge} \rangle + \langle \hat{\sigma}_i^{ee} \hat{\sigma}_j^{eg} \hat{\sigma}_n^{ge} \rangle  \right) \bigg\}, \nonumber \\
\end{align}  
\end{subequations}
\end{widetext}
\normalsize
where $(a \leftrightarrow b)$ indicates that an additional term appears equal to the previous one but with indexes $a$ and $b$ swapped. 

\section{Critical fraction of excited atoms}
\label{app: critical_fraction}
Let us consider an ensemble of N atoms. At initial timex, $N_\mathrm{exc}$ of them are excited, while the remaining $N_\mathrm{de}=N-N_\mathrm{exc}$ are de-excited (see Section~\ref{subsection: superradiance_partially_inverted}). We label as $\mathcal{E}$ and $\mathcal{D}$ the set of excited and de-excited atoms, respectively. We consider incoherently excited arrays, such that $\langle \hat{\sigma}_i^{ee} \rangle_0 = 1$ if $i \in \mathcal{E}$ and zero otherwise, and $ \langle \hat{\sigma}_i^{ee} \hat{\sigma}_j \rangle_0 =1$ if $i,j \in \mathcal{E}$ and zero otherwise. The derivative of the total emission rate for any configuration, given in Eq.~(\ref{eq: condition_burst_general}), can be expressed as
\begin{align}
    \dot{\gamma}_{\mathrm{tot},0} &= - \sum_i \frac{d^2}{dt^2} \langle \hat{\sigma}_i^{ee} \rangle_0, \label{eq: app_condition_burst} \\
    - \frac{d^2}{dt^2} \langle \hat{\sigma}_i^{ee} \rangle_0 &= - \Gamma_{0}^2   \langle \hat{\sigma}_i^{ee} \rangle_0 \nonumber \\
    &+ \sum_{j \neq i}  \frac{\Gamma_{ij} \Gamma_{ji}}{2} \left( 4\langle \hat{\sigma}_i^{ee} \hat{\sigma}_j^{ee} \rangle_0 - \langle \hat{\sigma}_i^{ee} \rangle_0 + \langle \hat{\sigma}_j^{ee} \rangle_0 \right). 
\end{align}
If all atoms are excited, one simply obtains $-\frac{d^2}{dt^2} \langle \hat{\sigma}_i^{ee} \rangle_0 = - \Gamma_{0}^2  + \sum_{j \neq i}  \Gamma_{ij} \Gamma_{ji}$. In the presence of de-excited atoms, one gets instead
\begin{align}
    - \frac{d^2}{dt^2} \langle \hat{\sigma}_{i \in \mathcal{E}} ^{ee} \rangle_0 &= - \Gamma_0^2 +  \sum_{j \neq i} \Gamma_{ij} \Gamma_{ji} - 1.5 \sum_{j \in D}  \Gamma_{ij} \Gamma_{ji} ,    \\ 
    - \frac{d^2}{dt^2} \langle \hat{\sigma}_{i \in \mathcal{D}} ^{ee} \rangle_0 &=   -0.5\sum_{j \in \mathcal{E}} \Gamma_{ij} \Gamma_{ji}  .
\end{align}
Plugging these expressions into Eq.(\ref{eq: app_condition_burst}), we obtain
\begin{align}
\label{eq: app_fraction_deexcited_oneconfig}
\dot{\gamma}_{\mathrm{tot},0} = &- \Gamma_0^2 N_\mathrm{exc} + \sum_i \sum_{j \neq i} \Gamma_{ij} \Gamma_{ji} \nonumber \\ &- 3 \sum_{i \in \mathcal{D}} \sum_{j \neq i} \Gamma_{ij} \Gamma_{ji} + 2 \sum_{i \in \mathcal{D}} \sum_{j \in D, j\neq i}\Gamma_{ij} \Gamma_{ji}.
\end{align}

We further average over all possible configurations of $N_\mathrm{de}$ de-excited atoms. We label the de-excited atoms with indexes $\alpha_1$ to $\alpha_{N_\mathrm{de}}$, each of them taking values from $1$ to $N$. Noting that no two indexes can be equal, the total number of permutations is $N !/(N-N_\mathrm{de})!$. Then, the average derivative of the total decay rate can be written as
\begin{equation}
    \bar{\dot{\gamma}}_\mathrm{tot,0} = - \Gamma_0^2 N_\mathrm{exc} + \sum_i \sum_{j \neq i} \Gamma_{ij} \Gamma_{ji} -3 \Sigma_1 + 2 \Sigma_2,
\end{equation}
where $\Sigma_1$ and $\Sigma_2$ are
\begin{widetext}
\small
\begin{align}
    \Sigma_1 &= \frac{(N-N_\mathrm{de})!}{N!} \sum_{\alpha_1} \sum_{\alpha_2 \neq \alpha_1} \ldots \sum_{\alpha_{N_{de}} \neq \alpha_1,\ldots,\alpha_{N_{de}-1}} \sum_{i \in \mathcal{D}} \sum_{j \neq i} \Gamma_{ij} \Gamma_{ji} \nonumber \\
    &= \frac{(N-N_\mathrm{de})!}{N!} \sum_{\alpha_1} \sum_{\alpha_2 \neq \alpha_1} \ldots \sum_{\alpha_{N_{de}} \neq \alpha_1,\ldots,\alpha_{N_{de}-1}} \left( \sum_{j \neq \alpha_1} \Gamma_{\alpha_1j} \Gamma_{j\alpha_1}  +\ldots+ \sum_{j \neq \alpha_{N_\mathrm{de}}} \Gamma_{\alpha_{N_\mathrm{de}}j} \Gamma_{j\alpha_{N_\mathrm{de}}} \right) \nonumber \\
    &= N_\mathrm{de} \frac{(N-N_\mathrm{de})!}{N!} \sum_{\alpha_1} \sum_{j \neq \alpha_1} \Gamma_{\alpha_1j} \Gamma_{j\alpha_1} \left( \sum_{\alpha_2 \neq \alpha_1} \ldots \sum_{\alpha_{N_{de}}\neq \alpha_1,\ldots,\alpha_{N_{de}-1}} \right) \nonumber \\
    &= \frac{N_\mathrm{de}}{N} \sum_i \sum_{j \neq i} \Gamma_{ij} \Gamma_{ji}, \label{eq: app_Sigma1} \\
    \Sigma_2 &= \frac{(N-N_\mathrm{de})!}{N!} \sum_{\alpha_1} \sum_{\alpha_2 \neq \alpha_1} \ldots \sum_{\alpha_{N_{de}} \neq \alpha_1,\ldots,\alpha_{N_{de}-1}} \sum_{i \in \mathcal{D}} \sum_{j \in \mathcal{D}, j\neq i} \Gamma_{ij} \Gamma_{ji} \nonumber \\
    &= \frac{(N-N_\mathrm{de})!}{N!} \sum_{\alpha_1} \sum_{\alpha_2 \neq \alpha_1} \ldots \sum_{\alpha_{N_{de}} \neq \alpha_1,\ldots,\alpha_{N_{de}-1}} \left( \Gamma_{\alpha_1 \alpha_2} \Gamma_{\alpha_2\alpha_1} +\ldots + \Gamma_{\alpha_1 \alpha_{N_\mathrm{de}}} \Gamma_{\alpha_{N_\mathrm{de}}\alpha_1} + \Gamma_{\alpha_2 \alpha_3} \Gamma_{\alpha_3\alpha_2} + \ldots  \right ) \nonumber \\
    &= N_\mathrm{de}(N_\mathrm{de}-1) \frac{(N-N_\mathrm{de})!}{N!} \sum_{\alpha_1} \sum_{\alpha_2 \neq \alpha_1} \Gamma_{\alpha_1 \alpha_2} \Gamma_{\alpha_2\alpha_1} \left( \sum_{\alpha_3 \neq \alpha_1,\alpha_2} \ldots \sum_{\alpha_{N_{de}}\neq \alpha_1,\ldots,\alpha_{N_{de}-1}} \right) \nonumber \\
    &= \frac{N_\mathrm{de}(N_\mathrm{de}-1)}{N(N-1)} \sum_i \sum_{j \neq i} \Gamma_{ij} \Gamma_{ji}, \label{eq: app_Sigma2}
\end{align}
\end{widetext}
\normalsize
and finally result in Eq.~(\ref{eq: deriv_tot_em_deexcited}) of the main text. 

\section{Critical filling fraction}
\label{app: critical_filling_fraction}
Let us consider an array with $N$ lattice position, $N_\mathrm{filled}$ of which are occupied and $N_\mathrm{hol} = N - N_\mathrm{filled}$ are empty (see Section~\ref{subsection: superradiance_missing_atoms}). Labelling $\mathcal{F}$ and $\mathcal{H}$ as the set of filled and empty lattice sites, we can compute the derivative of the total emission rate in a similar manner as in Appendix~\ref{app: critical_fraction}. Now, the second order derivatives of the population in lattice site $i$ read
\begin{align}
        - \frac{d^2}{dt^2} \langle \hat{\sigma}_{i \in \mathcal{F}} ^{ee} \rangle_0 &= - \Gamma_0^2 +  \sum_{j \in \mathcal{F}, j \neq i} \Gamma_{ij} \Gamma_{ji} ,    \\ 
    - \frac{d^2}{dt^2} \langle \hat{\sigma}_{i \in \mathcal{H}} ^{ee} \rangle_0 &=   0.
\end{align}
Note that the population at the lattice site of a missing atom is always zero, and so is its second derivative too. The population in a filled lattice site, on the other hand, is equal to that of an alternative, fully inverted system containing only $N_\mathrm{filled}$ atoms at the occupied positions. 

Then, $\dot{\gamma}_{\mathrm{tot},0}$ can be written as
\begin{align}
    \bar{\dot{\gamma}}_{\mathrm{tot},0} &= - \Gamma_0^2 N_\mathrm{filled} + \sum_{i \in \mathcal{F}} \sum_{j \in \mathcal{F}, j \neq i} \Gamma_{ij} \Gamma_{ji} \nonumber \\ 
    = &- \Gamma_0^2 N_\mathrm{filled} + \sum_i \sum_{j \neq i} \Gamma_{ij} \Gamma_{ji} \nonumber \\
    &- 2 \sum_{i \in \mathcal{H}} \sum_{j \neq i} \Gamma_{ij} \Gamma_{ji} + \sum_{i \in \mathcal{H}} \sum_{j \in H, j\neq i}\Gamma_{ij} \Gamma_{ji}.
\end{align}
which takes the same form as Eq.~(\ref{eq: app_fraction_deexcited_oneconfig}) but with different pre-factors. Using Eq.~(\ref{eq: app_Sigma1}) and Eq.~(\ref{eq: app_Sigma2}), one readily finds the average derivative of the total emission rate, given in Eq.~(\ref{eq: deriv_tot_em_holes}) of the main text.

\bibliographystyle{apsrev4-1-title}
\bibliography{reference_PRA_cumulants}

\begin{thebibliography}{57}%
\makeatletter
\providecommand \@ifxundefined [1]{%
 \@ifx{#1\undefined}
}%
\providecommand \@ifnum [1]{%
 \ifnum #1\expandafter \@firstoftwo
 \else \expandafter \@secondoftwo
 \fi
}%
\providecommand \@ifx [1]{%
 \ifx #1\expandafter \@firstoftwo
 \else \expandafter \@secondoftwo
 \fi
}%
\providecommand \natexlab [1]{#1}%
\providecommand \enquote  [1]{#1}%
\providecommand \bibnamefont  [1]{#1}%
\providecommand \bibfnamefont [1]{#1}%
\providecommand \citenamefont [1]{#1}%
\providecommand \href@noop [0]{\@secondoftwo}%
\providecommand \href [0]{\begingroup \@sanitize@url \@href}%
\providecommand \@href[1]{\@@startlink{#1}\@@href}%
\providecommand \@@href[1]{\endgroup#1\@@endlink}%
\providecommand \@sanitize@url [0]{\catcode `\\12\catcode `\$12\catcode
  `\&12\catcode `\#12\catcode `\^12\catcode `\_12\catcode `\%12\relax}%
\providecommand \@@startlink[1]{}%
\providecommand \@@endlink[0]{}%
\providecommand \url  [0]{\begingroup\@sanitize@url \@url }%
\providecommand \@url [1]{\endgroup\@href {#1}{\urlprefix }}%
\providecommand \urlprefix  [0]{URL }%
\providecommand \Eprint [0]{\href }%
\providecommand \doibase [0]{http://dx.doi.org/}%
\providecommand \selectlanguage [0]{\@gobble}%
\providecommand \bibinfo  [0]{\@secondoftwo}%
\providecommand \bibfield  [0]{\@secondoftwo}%
\providecommand \translation [1]{[#1]}%
\providecommand \BibitemOpen [0]{}%
\providecommand \bibitemStop [0]{}%
\providecommand \bibitemNoStop [0]{.\EOS\space}%
\providecommand \EOS [0]{\spacefactor3000\relax}%
\providecommand \BibitemShut  [1]{\csname bibitem#1\endcsname}%
\let\auto@bib@innerbib\@empty
\bibitem [{\citenamefont {Dicke}(1954)}]{Dicke_originalpaper}%
  \BibitemOpen
  \bibfield  {author} {\bibinfo {author} {\bibfnamefont {R.~H.}\ \bibnamefont
  {Dicke}},\ }\enquote {\bibinfo {title} {Coherence in Spontaneous Radiation
  Processes},}\ \href {\doibase 10.1103/PhysRev.93.99} {\bibfield  {journal}
  {\bibinfo  {journal} {Phys. Rev.}\ }\textbf {\bibinfo {volume} {93}},\
  \bibinfo {pages} {99} (\bibinfo {year} {1954})}\BibitemShut {NoStop}%
\bibitem [{\citenamefont {Gross}\ and\ \citenamefont
  {Haroche}(1982)}]{gross_haroche}%
  \BibitemOpen
  \bibfield  {author} {\bibinfo {author} {\bibfnamefont {M.}~\bibnamefont
  {Gross}}\ and\ \bibinfo {author} {\bibfnamefont {S.}~\bibnamefont
  {Haroche}},\ }\enquote {\bibinfo {title} {Superradiance: An essay on the
  theory of collective spontaneous emission},}\ \href {\doibase
  https://doi.org/10.1016/0370-1573(82)90102-8} {\bibfield  {journal} {\bibinfo
   {journal} {Physics Reports}\ }\textbf {\bibinfo {volume} {93}},\ \bibinfo
  {pages} {301} (\bibinfo {year} {1982})}\BibitemShut {NoStop}%
\bibitem [{\citenamefont {Rehler}\ and\ \citenamefont
  {Eberly}(1971)}]{Super_theory_Rehler}%
  \BibitemOpen
  \bibfield  {author} {\bibinfo {author} {\bibfnamefont {N.~E.}\ \bibnamefont
  {Rehler}}\ and\ \bibinfo {author} {\bibfnamefont {J.~H.}\ \bibnamefont
  {Eberly}},\ }\enquote {\bibinfo {title} {Superradiance},}\ \href {\doibase
  10.1103/PhysRevA.3.1735} {\bibfield  {journal} {\bibinfo  {journal} {Phys.
  Rev. A}\ }\textbf {\bibinfo {volume} {3}},\ \bibinfo {pages} {1735} (\bibinfo
  {year} {1971})}\BibitemShut {NoStop}%
\bibitem [{\citenamefont {Banfi}\ and\ \citenamefont
  {Bonifacio}(1975)}]{super_dicke_bonifacio}%
  \BibitemOpen
  \bibfield  {author} {\bibinfo {author} {\bibfnamefont {G.}~\bibnamefont
  {Banfi}}\ and\ \bibinfo {author} {\bibfnamefont {R.}~\bibnamefont
  {Bonifacio}},\ }\enquote {\bibinfo {title} {Superfluorescence and cooperative
  frequency shift},}\ \href {\doibase 10.1103/PhysRevA.12.2068} {\bibfield
  {journal} {\bibinfo  {journal} {Phys. Rev. A}\ }\textbf {\bibinfo {volume}
  {12}},\ \bibinfo {pages} {2068} (\bibinfo {year} {1975})}\BibitemShut
  {NoStop}%
\bibitem [{\citenamefont {Fuchs}\ \emph {et~al.}(2021)\citenamefont {Fuchs},
  \citenamefont {Vukics},\ and\ \citenamefont {Buhmann}}]{super_dicke_buhmann}%
  \BibitemOpen
  \bibfield  {author} {\bibinfo {author} {\bibfnamefont {S.}~\bibnamefont
  {Fuchs}}, \bibinfo {author} {\bibfnamefont {A.}~\bibnamefont {Vukics}}, \
  and\ \bibinfo {author} {\bibfnamefont {S.~Y.}\ \bibnamefont {Buhmann}},\
  }\enquote {\bibinfo {title} {Superradiance from nonideal initial states: A
  quantum trajectory approach},}\ \href {\doibase 10.1103/PhysRevA.103.043712}
  {\bibfield  {journal} {\bibinfo  {journal} {Phys. Rev. A}\ }\textbf {\bibinfo
  {volume} {103}},\ \bibinfo {pages} {043712} (\bibinfo {year}
  {2021})}\BibitemShut {NoStop}%
\bibitem [{\citenamefont {Shammah}\ \emph {et~al.}(2017)\citenamefont
  {Shammah}, \citenamefont {Lambert}, \citenamefont {Nori},\ and\ \citenamefont
  {De~Liberato}}]{super_dicke_liberato}%
  \BibitemOpen
  \bibfield  {author} {\bibinfo {author} {\bibfnamefont {N.}~\bibnamefont
  {Shammah}}, \bibinfo {author} {\bibfnamefont {N.}~\bibnamefont {Lambert}},
  \bibinfo {author} {\bibfnamefont {F.}~\bibnamefont {Nori}}, \ and\ \bibinfo
  {author} {\bibfnamefont {S.}~\bibnamefont {De~Liberato}},\ }\enquote
  {\bibinfo {title} {Superradiance with local phase-breaking effects},}\ \href
  {\doibase 10.1103/PhysRevA.96.023863} {\bibfield  {journal} {\bibinfo
  {journal} {Phys. Rev. A}\ }\textbf {\bibinfo {volume} {96}},\ \bibinfo
  {pages} {023863} (\bibinfo {year} {2017})}\BibitemShut {NoStop}%
\bibitem [{\citenamefont {Malz}\ \emph {et~al.}(2022)\citenamefont {Malz},
  \citenamefont {Trivedi},\ and\ \citenamefont {Cirac}}]{super_dicke_malz}%
  \BibitemOpen
  \bibfield  {author} {\bibinfo {author} {\bibfnamefont {D.}~\bibnamefont
  {Malz}}, \bibinfo {author} {\bibfnamefont {R.}~\bibnamefont {Trivedi}}, \
  and\ \bibinfo {author} {\bibfnamefont {J.~I.}\ \bibnamefont {Cirac}},\
  }\enquote {\bibinfo {title} {Large-$N$ limit of Dicke superradiance},}\ \href
  {\doibase 10.1103/PhysRevA.106.013716} {\bibfield  {journal} {\bibinfo
  {journal} {Phys. Rev. A}\ }\textbf {\bibinfo {volume} {106}},\ \bibinfo
  {pages} {013716} (\bibinfo {year} {2022})}\BibitemShut {NoStop}%
\bibitem [{\citenamefont {Lemberger}\ and\ \citenamefont
  {M\o{}lmer}(2021)}]{super_dicke_molmer}%
  \BibitemOpen
  \bibfield  {author} {\bibinfo {author} {\bibfnamefont {B.}~\bibnamefont
  {Lemberger}}\ and\ \bibinfo {author} {\bibfnamefont {K.}~\bibnamefont
  {M\o{}lmer}},\ }\enquote {\bibinfo {title} {Radiation eigenmodes of Dicke
  superradiance},}\ \href {\doibase 10.1103/PhysRevA.103.033713} {\bibfield
  {journal} {\bibinfo  {journal} {Phys. Rev. A}\ }\textbf {\bibinfo {volume}
  {103}},\ \bibinfo {pages} {033713} (\bibinfo {year} {2021})}\BibitemShut
  {NoStop}%
\bibitem [{\citenamefont {Inouye}\ \emph {et~al.}(1999)\citenamefont {Inouye},
  \citenamefont {Chikkatur}, \citenamefont {Stamper-Kurn}, \citenamefont
  {Stenger}, \citenamefont {Pritchard},\ and\ \citenamefont
  {Ketterle}}]{exp_super_atoms_1}%
  \BibitemOpen
  \bibfield  {author} {\bibinfo {author} {\bibfnamefont {S.}~\bibnamefont
  {Inouye}}, \bibinfo {author} {\bibfnamefont {A.~P.}\ \bibnamefont
  {Chikkatur}}, \bibinfo {author} {\bibfnamefont {D.~M.}\ \bibnamefont
  {Stamper-Kurn}}, \bibinfo {author} {\bibfnamefont {J.}~\bibnamefont
  {Stenger}}, \bibinfo {author} {\bibfnamefont {D.~E.}\ \bibnamefont
  {Pritchard}}, \ and\ \bibinfo {author} {\bibfnamefont {W.}~\bibnamefont
  {Ketterle}},\ }\enquote {\bibinfo {title} {Superradiant Rayleigh Scattering
  from a Bose-Einstein Condensate},}\ \href {\doibase
  10.1126/science.285.5427.571} {\bibfield  {journal} {\bibinfo  {journal}
  {Science}\ }\textbf {\bibinfo {volume} {285}},\ \bibinfo {pages} {571}
  (\bibinfo {year} {1999})}\BibitemShut {NoStop}%
\bibitem [{\citenamefont {Wang}\ \emph {et~al.}(2007)\citenamefont {Wang},
  \citenamefont {Yelin}, \citenamefont {C\^ot\'e}, \citenamefont {Eyler},
  \citenamefont {Farooqi}, \citenamefont {Gould}, \citenamefont
  {Ko\ifmmode~\check{s}\else \v{s}\fi{}trun}, \citenamefont {Tong},\ and\
  \citenamefont {Vrinceanu}}]{exp_super_atoms_2}%
  \BibitemOpen
  \bibfield  {author} {\bibinfo {author} {\bibfnamefont {T.}~\bibnamefont
  {Wang}}, \bibinfo {author} {\bibfnamefont {S.~F.}\ \bibnamefont {Yelin}},
  \bibinfo {author} {\bibfnamefont {R.}~\bibnamefont {C\^ot\'e}}, \bibinfo
  {author} {\bibfnamefont {E.~E.}\ \bibnamefont {Eyler}}, \bibinfo {author}
  {\bibfnamefont {S.~M.}\ \bibnamefont {Farooqi}}, \bibinfo {author}
  {\bibfnamefont {P.~L.}\ \bibnamefont {Gould}}, \bibinfo {author}
  {\bibfnamefont {M.}~\bibnamefont {Ko\ifmmode~\check{s}\else \v{s}\fi{}trun}},
  \bibinfo {author} {\bibfnamefont {D.}~\bibnamefont {Tong}}, \ and\ \bibinfo
  {author} {\bibfnamefont {D.}~\bibnamefont {Vrinceanu}},\ }\enquote {\bibinfo
  {title} {Superradiance in ultracold Rydberg gases},}\ \href {\doibase
  10.1103/PhysRevA.75.033802} {\bibfield  {journal} {\bibinfo  {journal} {Phys.
  Rev. A}\ }\textbf {\bibinfo {volume} {75}},\ \bibinfo {pages} {033802}
  (\bibinfo {year} {2007})}\BibitemShut {NoStop}%
\bibitem [{\citenamefont {Grimes}\ \emph {et~al.}(2017)\citenamefont {Grimes},
  \citenamefont {Coy}, \citenamefont {Barnum}, \citenamefont {Zhou},
  \citenamefont {Yelin},\ and\ \citenamefont {Field}}]{exp_super_atoms_3}%
  \BibitemOpen
  \bibfield  {author} {\bibinfo {author} {\bibfnamefont {D.~D.}\ \bibnamefont
  {Grimes}}, \bibinfo {author} {\bibfnamefont {S.~L.}\ \bibnamefont {Coy}},
  \bibinfo {author} {\bibfnamefont {T.~J.}\ \bibnamefont {Barnum}}, \bibinfo
  {author} {\bibfnamefont {Y.}~\bibnamefont {Zhou}}, \bibinfo {author}
  {\bibfnamefont {S.~F.}\ \bibnamefont {Yelin}}, \ and\ \bibinfo {author}
  {\bibfnamefont {R.~W.}\ \bibnamefont {Field}},\ }\enquote {\bibinfo {title}
  {Direct single-shot observation of millimeter-wave superradiance in
  Rydberg-Rydberg transitions},}\ \href {\doibase 10.1103/PhysRevA.95.043818}
  {\bibfield  {journal} {\bibinfo  {journal} {Phys. Rev. A}\ }\textbf {\bibinfo
  {volume} {95}},\ \bibinfo {pages} {043818} (\bibinfo {year}
  {2017})}\BibitemShut {NoStop}%
\bibitem [{\citenamefont {Kaluzny}\ \emph {et~al.}(1983)\citenamefont
  {Kaluzny}, \citenamefont {Goy}, \citenamefont {Gross}, \citenamefont
  {Raimond},\ and\ \citenamefont {Haroche}}]{exp_super_atoms_4}%
  \BibitemOpen
  \bibfield  {author} {\bibinfo {author} {\bibfnamefont {Y.}~\bibnamefont
  {Kaluzny}}, \bibinfo {author} {\bibfnamefont {P.}~\bibnamefont {Goy}},
  \bibinfo {author} {\bibfnamefont {M.}~\bibnamefont {Gross}}, \bibinfo
  {author} {\bibfnamefont {J.~M.}\ \bibnamefont {Raimond}}, \ and\ \bibinfo
  {author} {\bibfnamefont {S.}~\bibnamefont {Haroche}},\ }\enquote {\bibinfo
  {title} {Observation of Self-Induced Rabi Oscillations in Two-Level Atoms
  Excited Inside a Resonant Cavity: The Ringing Regime of Superradiance},}\
  \href {\doibase 10.1103/PhysRevLett.51.1175} {\bibfield  {journal} {\bibinfo
  {journal} {Phys. Rev. Lett.}\ }\textbf {\bibinfo {volume} {51}},\ \bibinfo
  {pages} {1175} (\bibinfo {year} {1983})}\BibitemShut {NoStop}%
\bibitem [{\citenamefont {Ara\'ujo}\ \emph {et~al.}(2016)\citenamefont
  {Ara\'ujo}, \citenamefont {Kre\ifmmode \check{s}\else
  \v{s}\fi{}i\ifmmode~\acute{c}\else \'{c}\fi{}}, \citenamefont {Kaiser},\ and\
  \citenamefont {Guerin}}]{exp_super_atoms_5}%
  \BibitemOpen
  \bibfield  {author} {\bibinfo {author} {\bibfnamefont {M.~O.}\ \bibnamefont
  {Ara\'ujo}}, \bibinfo {author} {\bibfnamefont {I.}~\bibnamefont {Kre\ifmmode
  \check{s}\else \v{s}\fi{}i\ifmmode~\acute{c}\else \'{c}\fi{}}}, \bibinfo
  {author} {\bibfnamefont {R.}~\bibnamefont {Kaiser}}, \ and\ \bibinfo {author}
  {\bibfnamefont {W.}~\bibnamefont {Guerin}},\ }\enquote {\bibinfo {title}
  {Superradiance in a Large and Dilute Cloud of Cold Atoms in the Linear-Optics
  Regime},}\ \href {\doibase 10.1103/PhysRevLett.117.073002} {\bibfield
  {journal} {\bibinfo  {journal} {Phys. Rev. Lett.}\ }\textbf {\bibinfo
  {volume} {117}},\ \bibinfo {pages} {073002} (\bibinfo {year}
  {2016})}\BibitemShut {NoStop}%
\bibitem [{\citenamefont {Chen}\ \emph {et~al.}(2018)\citenamefont {Chen},
  \citenamefont {Wang}, \citenamefont {Meng}, \citenamefont {Huang},
  \citenamefont {Cai}, \citenamefont {Wang}, \citenamefont {Zhu},\ and\
  \citenamefont {Zhang}}]{exp_super_atoms_6}%
  \BibitemOpen
  \bibfield  {author} {\bibinfo {author} {\bibfnamefont {L.}~\bibnamefont
  {Chen}}, \bibinfo {author} {\bibfnamefont {P.}~\bibnamefont {Wang}}, \bibinfo
  {author} {\bibfnamefont {Z.}~\bibnamefont {Meng}}, \bibinfo {author}
  {\bibfnamefont {L.}~\bibnamefont {Huang}}, \bibinfo {author} {\bibfnamefont
  {H.}~\bibnamefont {Cai}}, \bibinfo {author} {\bibfnamefont {D.-W.}\
  \bibnamefont {Wang}}, \bibinfo {author} {\bibfnamefont {S.-Y.}\ \bibnamefont
  {Zhu}}, \ and\ \bibinfo {author} {\bibfnamefont {J.}~\bibnamefont {Zhang}},\
  }\enquote {\bibinfo {title} {Experimental Observation of One-Dimensional
  Superradiance Lattices in Ultracold Atoms},}\ \href {\doibase
  10.1103/PhysRevLett.120.193601} {\bibfield  {journal} {\bibinfo  {journal}
  {Phys. Rev. Lett.}\ }\textbf {\bibinfo {volume} {120}},\ \bibinfo {pages}
  {193601} (\bibinfo {year} {2018})}\BibitemShut {NoStop}%
\bibitem [{\citenamefont {Goban}\ \emph {et~al.}(2015)\citenamefont {Goban},
  \citenamefont {Hung}, \citenamefont {Hood}, \citenamefont {Yu}, \citenamefont
  {Muniz}, \citenamefont {Painter},\ and\ \citenamefont
  {Kimble}}]{exp_super_atoms_7}%
  \BibitemOpen
  \bibfield  {author} {\bibinfo {author} {\bibfnamefont {A.}~\bibnamefont
  {Goban}}, \bibinfo {author} {\bibfnamefont {C.-L.}\ \bibnamefont {Hung}},
  \bibinfo {author} {\bibfnamefont {J.~D.}\ \bibnamefont {Hood}}, \bibinfo
  {author} {\bibfnamefont {S.-P.}\ \bibnamefont {Yu}}, \bibinfo {author}
  {\bibfnamefont {J.~A.}\ \bibnamefont {Muniz}}, \bibinfo {author}
  {\bibfnamefont {O.}~\bibnamefont {Painter}}, \ and\ \bibinfo {author}
  {\bibfnamefont {H.~J.}\ \bibnamefont {Kimble}},\ }\enquote {\bibinfo {title}
  {Superradiance for Atoms Trapped along a Photonic Crystal Waveguide},}\ \href
  {\doibase 10.1103/PhysRevLett.115.063601} {\bibfield  {journal} {\bibinfo
  {journal} {Phys. Rev. Lett.}\ }\textbf {\bibinfo {volume} {115}},\ \bibinfo
  {pages} {063601} (\bibinfo {year} {2015})}\BibitemShut {NoStop}%
\bibitem [{\citenamefont {Scheibner}\ \emph {et~al.}(2007)\citenamefont
  {Scheibner}, \citenamefont {Schmidt}, \citenamefont {Worschech},
  \citenamefont {Forchel}, \citenamefont {Bacher}, \citenamefont {Passow},\
  and\ \citenamefont {Hommel}}]{experiment_quantumdots_1}%
  \BibitemOpen
  \bibfield  {author} {\bibinfo {author} {\bibfnamefont {M.}~\bibnamefont
  {Scheibner}}, \bibinfo {author} {\bibfnamefont {T.}~\bibnamefont {Schmidt}},
  \bibinfo {author} {\bibfnamefont {L.}~\bibnamefont {Worschech}}, \bibinfo
  {author} {\bibfnamefont {A.}~\bibnamefont {Forchel}}, \bibinfo {author}
  {\bibfnamefont {G.}~\bibnamefont {Bacher}}, \bibinfo {author} {\bibfnamefont
  {T.}~\bibnamefont {Passow}}, \ and\ \bibinfo {author} {\bibfnamefont
  {D.}~\bibnamefont {Hommel}},\ }\enquote {\bibinfo {title} {Superradiance of
  quantum dots},}\ \href {\doibase 10.1038/nphys494} {\bibfield  {journal}
  {\bibinfo  {journal} {Nature Physics}\ }\textbf {\bibinfo {volume} {3}},\
  \bibinfo {pages} {106} (\bibinfo {year} {2007})}\BibitemShut {NoStop}%
\bibitem [{\citenamefont {Rain{\`o}}\ \emph {et~al.}(2018)\citenamefont
  {Rain{\`o}}, \citenamefont {Becker}, \citenamefont {Bodnarchuk},
  \citenamefont {Mahrt}, \citenamefont {Kovalenko},\ and\ \citenamefont
  {St{\"o}ferle}}]{experiment_quantumdots_2}%
  \BibitemOpen
  \bibfield  {author} {\bibinfo {author} {\bibfnamefont {G.}~\bibnamefont
  {Rain{\`o}}}, \bibinfo {author} {\bibfnamefont {M.~A.}\ \bibnamefont
  {Becker}}, \bibinfo {author} {\bibfnamefont {M.~I.}\ \bibnamefont
  {Bodnarchuk}}, \bibinfo {author} {\bibfnamefont {R.~F.}\ \bibnamefont
  {Mahrt}}, \bibinfo {author} {\bibfnamefont {M.~V.}\ \bibnamefont
  {Kovalenko}}, \ and\ \bibinfo {author} {\bibfnamefont {T.}~\bibnamefont
  {St{\"o}ferle}},\ }\enquote {\bibinfo {title} {Superfluorescence from lead
  halide perovskite quantum dot superlattices},}\ \href {\doibase
  10.1038/s41586-018-0683-0} {\bibfield  {journal} {\bibinfo  {journal}
  {Nature}\ }\textbf {\bibinfo {volume} {563}},\ \bibinfo {pages} {671}
  (\bibinfo {year} {2018})}\BibitemShut {NoStop}%
\bibitem [{\citenamefont {Bradac}\ \emph {et~al.}(2017)\citenamefont {Bradac},
  \citenamefont {Johnsson}, \citenamefont {Breugel}, \citenamefont {Baragiola},
  \citenamefont {Martin}, \citenamefont {Juan}, \citenamefont {Brennen},\ and\
  \citenamefont {Volz}}]{experiment_NV}%
  \BibitemOpen
  \bibfield  {author} {\bibinfo {author} {\bibfnamefont {C.}~\bibnamefont
  {Bradac}}, \bibinfo {author} {\bibfnamefont {M.~T.}\ \bibnamefont
  {Johnsson}}, \bibinfo {author} {\bibfnamefont {M.~v.}\ \bibnamefont
  {Breugel}}, \bibinfo {author} {\bibfnamefont {B.~Q.}\ \bibnamefont
  {Baragiola}}, \bibinfo {author} {\bibfnamefont {R.}~\bibnamefont {Martin}},
  \bibinfo {author} {\bibfnamefont {M.~L.}\ \bibnamefont {Juan}}, \bibinfo
  {author} {\bibfnamefont {G.~K.}\ \bibnamefont {Brennen}}, \ and\ \bibinfo
  {author} {\bibfnamefont {T.}~\bibnamefont {Volz}},\ }\enquote {\bibinfo
  {title} {Room-temperature spontaneous superradiance from single diamond
  nanocrystals},}\ \href {\doibase 10.1038/s41467-017-01397-4} {\bibfield
  {journal} {\bibinfo  {journal} {Nature Communications}\ }\textbf {\bibinfo
  {volume} {8}},\ \bibinfo {pages} {1205} (\bibinfo {year} {2017})}\BibitemShut
  {NoStop}%
\bibitem [{\citenamefont {Haider}\ \emph {et~al.}(2021)\citenamefont {Haider},
  \citenamefont {Sampathkumar}, \citenamefont {Verhagen}, \citenamefont
  {Nádvorník}, \citenamefont {Sonia}, \citenamefont {Valeš}, \citenamefont
  {Sýkora}, \citenamefont {Kapusta}, \citenamefont {Němec}, \citenamefont
  {Hof}, \citenamefont {Frank}, \citenamefont {Chen}, \citenamefont
  {Vejpravová},\ and\ \citenamefont {Kalbáč}}]{experiment_2dmat}%
  \BibitemOpen
  \bibfield  {author} {\bibinfo {author} {\bibfnamefont {G.}~\bibnamefont
  {Haider}}, \bibinfo {author} {\bibfnamefont {K.}~\bibnamefont
  {Sampathkumar}}, \bibinfo {author} {\bibfnamefont {T.}~\bibnamefont
  {Verhagen}}, \bibinfo {author} {\bibfnamefont {L.}~\bibnamefont
  {Nádvorník}}, \bibinfo {author} {\bibfnamefont {F.~J.}\ \bibnamefont
  {Sonia}}, \bibinfo {author} {\bibfnamefont {V.}~\bibnamefont {Valeš}},
  \bibinfo {author} {\bibfnamefont {J.}~\bibnamefont {Sýkora}}, \bibinfo
  {author} {\bibfnamefont {P.}~\bibnamefont {Kapusta}}, \bibinfo {author}
  {\bibfnamefont {P.}~\bibnamefont {Němec}}, \bibinfo {author} {\bibfnamefont
  {M.}~\bibnamefont {Hof}}, \bibinfo {author} {\bibfnamefont {O.}~\bibnamefont
  {Frank}}, \bibinfo {author} {\bibfnamefont {Y.-F.}\ \bibnamefont {Chen}},
  \bibinfo {author} {\bibfnamefont {J.}~\bibnamefont {Vejpravová}}, \ and\
  \bibinfo {author} {\bibfnamefont {M.}~\bibnamefont {Kalbáč}},\ }\enquote
  {\bibinfo {title} {Superradiant Emission from Coherent Excitons in van Der
  Waals Heterostructures},}\ \href {\doibase 10.1002/adfm.202102196} {\bibfield
   {journal} {\bibinfo  {journal} {Advanced Functional Materials}\ }\textbf
  {\bibinfo {volume} {31}},\ \bibinfo {pages} {2102196} (\bibinfo {year}
  {2021})}\BibitemShut {NoStop}%
\bibitem [{\citenamefont {Asenjo-Garcia}\ \emph {et~al.}(2017)\citenamefont
  {Asenjo-Garcia}, \citenamefont {Moreno-Cardoner}, \citenamefont {Albrecht},
  \citenamefont {Kimble},\ and\ \citenamefont
  {Chang}}]{AnaAsenjo_2017_general}%
  \BibitemOpen
  \bibfield  {author} {\bibinfo {author} {\bibfnamefont {A.}~\bibnamefont
  {Asenjo-Garcia}}, \bibinfo {author} {\bibfnamefont {M.}~\bibnamefont
  {Moreno-Cardoner}}, \bibinfo {author} {\bibfnamefont {A.}~\bibnamefont
  {Albrecht}}, \bibinfo {author} {\bibfnamefont {H.~J.}\ \bibnamefont
  {Kimble}}, \ and\ \bibinfo {author} {\bibfnamefont {D.~E.}\ \bibnamefont
  {Chang}},\ }\enquote {\bibinfo {title} {Exponential Improvement in Photon
  Storage Fidelities Using Subradiance and ``Selective Radiance'' in Atomic
  Arrays},}\ \href {\doibase 10.1103/PhysRevX.7.031024} {\bibfield  {journal}
  {\bibinfo  {journal} {Phys. Rev. X}\ }\textbf {\bibinfo {volume} {7}},\
  \bibinfo {pages} {031024} (\bibinfo {year} {2017})}\BibitemShut {NoStop}%
\bibitem [{\citenamefont {Rui}\ \emph {et~al.}(2020)\citenamefont {Rui},
  \citenamefont {Wei}, \citenamefont {Rubio-Abadal}, \citenamefont {Hollerith},
  \citenamefont {Zeiher}, \citenamefont {Stamper-Kurn}, \citenamefont {Gross},\
  and\ \citenamefont {Bloch}}]{Bloch_mirror}%
  \BibitemOpen
  \bibfield  {author} {\bibinfo {author} {\bibfnamefont {J.}~\bibnamefont
  {Rui}}, \bibinfo {author} {\bibfnamefont {D.}~\bibnamefont {Wei}}, \bibinfo
  {author} {\bibfnamefont {A.}~\bibnamefont {Rubio-Abadal}}, \bibinfo {author}
  {\bibfnamefont {S.}~\bibnamefont {Hollerith}}, \bibinfo {author}
  {\bibfnamefont {J.}~\bibnamefont {Zeiher}}, \bibinfo {author} {\bibfnamefont
  {D.~M.}\ \bibnamefont {Stamper-Kurn}}, \bibinfo {author} {\bibfnamefont
  {C.}~\bibnamefont {Gross}}, \ and\ \bibinfo {author} {\bibfnamefont
  {I.}~\bibnamefont {Bloch}},\ }\enquote {\bibinfo {title} {A subradiant
  optical mirror formed by a single structured atomic layer},}\ \href {\doibase
  10.1038/s41586-020-2463-x} {\bibfield  {journal} {\bibinfo  {journal}
  {Nature}\ }\textbf {\bibinfo {volume} {583}},\ \bibinfo {pages} {369}
  (\bibinfo {year} {2020})}\BibitemShut {NoStop}%
\bibitem [{\citenamefont {Shahmoon}\ \emph {et~al.}(2017)\citenamefont
  {Shahmoon}, \citenamefont {Wild}, \citenamefont {Lukin},\ and\ \citenamefont
  {Yelin}}]{Ephi_2017}%
  \BibitemOpen
  \bibfield  {author} {\bibinfo {author} {\bibfnamefont {E.}~\bibnamefont
  {Shahmoon}}, \bibinfo {author} {\bibfnamefont {D.~S.}\ \bibnamefont {Wild}},
  \bibinfo {author} {\bibfnamefont {M.~D.}\ \bibnamefont {Lukin}}, \ and\
  \bibinfo {author} {\bibfnamefont {S.~F.}\ \bibnamefont {Yelin}},\ }\enquote
  {\bibinfo {title} {Cooperative Resonances in Light Scattering from
  Two-Dimensional Atomic Arrays},}\ \href {\doibase
  10.1103/PhysRevLett.118.113601} {\bibfield  {journal} {\bibinfo  {journal}
  {Phys. Rev. Lett.}\ }\textbf {\bibinfo {volume} {118}},\ \bibinfo {pages}
  {113601} (\bibinfo {year} {2017})}\BibitemShut {NoStop}%
\bibitem [{\citenamefont {Bettles}\ \emph {et~al.}(2016)\citenamefont
  {Bettles}, \citenamefont {Gardiner},\ and\ \citenamefont
  {Adams}}]{PRL_CAdams}%
  \BibitemOpen
  \bibfield  {author} {\bibinfo {author} {\bibfnamefont {R.~J.}\ \bibnamefont
  {Bettles}}, \bibinfo {author} {\bibfnamefont {S.~A.}\ \bibnamefont
  {Gardiner}}, \ and\ \bibinfo {author} {\bibfnamefont {C.~S.}\ \bibnamefont
  {Adams}},\ }\enquote {\bibinfo {title} {Enhanced Optical Cross Section via
  Collective Coupling of Atomic Dipoles in a 2D Array},}\ \href {\doibase
  10.1103/PhysRevLett.116.103602} {\bibfield  {journal} {\bibinfo  {journal}
  {Phys. Rev. Lett.}\ }\textbf {\bibinfo {volume} {116}},\ \bibinfo {pages}
  {103602} (\bibinfo {year} {2016})}\BibitemShut {NoStop}%
\bibitem [{\citenamefont {Rubies-Bigorda}\ \emph
  {et~al.}(2022{\natexlab{a}})\citenamefont {Rubies-Bigorda}, \citenamefont
  {Walther}, \citenamefont {Patti},\ and\ \citenamefont
  {Yelin}}]{Oriol_subradiance}%
  \BibitemOpen
  \bibfield  {author} {\bibinfo {author} {\bibfnamefont {O.}~\bibnamefont
  {Rubies-Bigorda}}, \bibinfo {author} {\bibfnamefont {V.}~\bibnamefont
  {Walther}}, \bibinfo {author} {\bibfnamefont {T.~L.}\ \bibnamefont {Patti}},
  \ and\ \bibinfo {author} {\bibfnamefont {S.~F.}\ \bibnamefont {Yelin}},\
  }\enquote {\bibinfo {title} {Photon control and coherent interactions via
  lattice dark states in atomic arrays},}\ \href {\doibase
  10.1103/PhysRevResearch.4.013110} {\bibfield  {journal} {\bibinfo  {journal}
  {Phys. Rev. Research}\ }\textbf {\bibinfo {volume} {4}},\ \bibinfo {pages}
  {013110} (\bibinfo {year} {2022}{\natexlab{a}})}\BibitemShut {NoStop}%
\bibitem [{\citenamefont {Ballantine}\ and\ \citenamefont
  {Ruostekoski}(2021)}]{Ruostekoski_subradiance_PRX}%
  \BibitemOpen
  \bibfield  {author} {\bibinfo {author} {\bibfnamefont {K.~E.}\ \bibnamefont
  {Ballantine}}\ and\ \bibinfo {author} {\bibfnamefont {J.}~\bibnamefont
  {Ruostekoski}},\ }\enquote {\bibinfo {title} {Quantum Single-Photon Control,
  Storage, and Entanglement Generation with Planar Atomic Arrays},}\ \href
  {\doibase 10.1103/PRXQuantum.2.040362} {\bibfield  {journal} {\bibinfo
  {journal} {PRX Quantum}\ }\textbf {\bibinfo {volume} {2}},\ \bibinfo {pages}
  {040362} (\bibinfo {year} {2021})}\BibitemShut {NoStop}%
\bibitem [{\citenamefont {Moreno-Cardoner}\ \emph {et~al.}(2022)\citenamefont
  {Moreno-Cardoner}, \citenamefont {Holzinger},\ and\ \citenamefont
  {Ritsch}}]{Ritsch_subradiance_9fold}%
  \BibitemOpen
  \bibfield  {author} {\bibinfo {author} {\bibfnamefont {M.}~\bibnamefont
  {Moreno-Cardoner}}, \bibinfo {author} {\bibfnamefont {R.}~\bibnamefont
  {Holzinger}}, \ and\ \bibinfo {author} {\bibfnamefont {H.}~\bibnamefont
  {Ritsch}},\ }\enquote {\bibinfo {title} {Efficient nano-photonic antennas
  based on dark states in quantum emitter rings},}\ \href {\doibase
  10.1364/OE.437396} {\bibfield  {journal} {\bibinfo  {journal} {Opt. Express}\
  }\textbf {\bibinfo {volume} {30}},\ \bibinfo {pages} {10779} (\bibinfo {year}
  {2022})}\BibitemShut {NoStop}%
\bibitem [{\citenamefont {Perczel}\ \emph {et~al.}(2017)\citenamefont
  {Perczel}, \citenamefont {Borregaard}, \citenamefont {Chang}, \citenamefont
  {Pichler}, \citenamefont {Yelin}, \citenamefont {Zoller},\ and\ \citenamefont
  {Lukin}}]{Janos_2017_PRL}%
  \BibitemOpen
  \bibfield  {author} {\bibinfo {author} {\bibfnamefont {J.}~\bibnamefont
  {Perczel}}, \bibinfo {author} {\bibfnamefont {J.}~\bibnamefont {Borregaard}},
  \bibinfo {author} {\bibfnamefont {D.~E.}\ \bibnamefont {Chang}}, \bibinfo
  {author} {\bibfnamefont {H.}~\bibnamefont {Pichler}}, \bibinfo {author}
  {\bibfnamefont {S.~F.}\ \bibnamefont {Yelin}}, \bibinfo {author}
  {\bibfnamefont {P.}~\bibnamefont {Zoller}}, \ and\ \bibinfo {author}
  {\bibfnamefont {M.~D.}\ \bibnamefont {Lukin}},\ }\enquote {\bibinfo {title}
  {Topological Quantum Optics in Two-Dimensional Atomic Arrays},}\ \href
  {\doibase 10.1103/PhysRevLett.119.023603} {\bibfield  {journal} {\bibinfo
  {journal} {Phys. Rev. Lett.}\ }\textbf {\bibinfo {volume} {119}},\ \bibinfo
  {pages} {023603} (\bibinfo {year} {2017})}\BibitemShut {NoStop}%
\bibitem [{\citenamefont {Shahmoon}\ \emph {et~al.}(2020)\citenamefont
  {Shahmoon}, \citenamefont {Lukin},\ and\ \citenamefont
  {Yelin}}]{PRA_motion_Ephi}%
  \BibitemOpen
  \bibfield  {author} {\bibinfo {author} {\bibfnamefont {E.}~\bibnamefont
  {Shahmoon}}, \bibinfo {author} {\bibfnamefont {M.~D.}\ \bibnamefont {Lukin}},
  \ and\ \bibinfo {author} {\bibfnamefont {S.~F.}\ \bibnamefont {Yelin}},\
  }\enquote {\bibinfo {title} {Quantum optomechanics of a two-dimensional
  atomic array},}\ \href {\doibase 10.1103/PhysRevA.101.063833} {\bibfield
  {journal} {\bibinfo  {journal} {Phys. Rev. A}\ }\textbf {\bibinfo {volume}
  {101}},\ \bibinfo {pages} {063833} (\bibinfo {year} {2020})}\BibitemShut
  {NoStop}%
\bibitem [{\citenamefont {Moreno-Cardoner}\ \emph {et~al.}(2021)\citenamefont
  {Moreno-Cardoner}, \citenamefont {Goncalves},\ and\ \citenamefont
  {Chang}}]{MorenoCardoner_Rydberg}%
  \BibitemOpen
  \bibfield  {author} {\bibinfo {author} {\bibfnamefont {M.}~\bibnamefont
  {Moreno-Cardoner}}, \bibinfo {author} {\bibfnamefont {D.}~\bibnamefont
  {Goncalves}}, \ and\ \bibinfo {author} {\bibfnamefont {D.~E.}\ \bibnamefont
  {Chang}},\ }\enquote {\bibinfo {title} {Quantum Nonlinear Optics Based on
  Two-Dimensional Rydberg Atom Arrays},}\ \href {\doibase
  10.1103/PhysRevLett.127.263602} {\bibfield  {journal} {\bibinfo  {journal}
  {Phys. Rev. Lett.}\ }\textbf {\bibinfo {volume} {127}},\ \bibinfo {pages}
  {263602} (\bibinfo {year} {2021})}\BibitemShut {NoStop}%
\bibitem [{\citenamefont {Buckley-Bonanno}\ \emph {et~al.}(2022)\citenamefont
  {Buckley-Bonanno}, \citenamefont {Ostermann}, \citenamefont {Rubies-Bigorda},
  \citenamefont {Patti},\ and\ \citenamefont {Yelin}}]{Samuel_PRA}%
  \BibitemOpen
  \bibfield  {author} {\bibinfo {author} {\bibfnamefont {S.}~\bibnamefont
  {Buckley-Bonanno}}, \bibinfo {author} {\bibfnamefont {S.}~\bibnamefont
  {Ostermann}}, \bibinfo {author} {\bibfnamefont {O.}~\bibnamefont
  {Rubies-Bigorda}}, \bibinfo {author} {\bibfnamefont {T.~L.}\ \bibnamefont
  {Patti}}, \ and\ \bibinfo {author} {\bibfnamefont {S.~F.}\ \bibnamefont
  {Yelin}},\ }\enquote {\bibinfo {title} {Optimized geometries for cooperative
  photon storage in an impurity coupled to a two-dimensional atomic array},}\
  \href {\doibase 10.1103/PhysRevA.106.053706} {\bibfield  {journal} {\bibinfo
  {journal} {Phys. Rev. A}\ }\textbf {\bibinfo {volume} {106}},\ \bibinfo
  {pages} {053706} (\bibinfo {year} {2022})}\BibitemShut {NoStop}%
\bibitem [{\citenamefont {Zhang}\ \emph {et~al.}(2022)\citenamefont {Zhang},
  \citenamefont {Walther}, \citenamefont {M{\o{}}lmer},\ and\ \citenamefont
  {Pohl}}]{Zhang2022photonphoton}%
  \BibitemOpen
  \bibfield  {author} {\bibinfo {author} {\bibfnamefont {L.}~\bibnamefont
  {Zhang}}, \bibinfo {author} {\bibfnamefont {V.}~\bibnamefont {Walther}},
  \bibinfo {author} {\bibfnamefont {K.}~\bibnamefont {M{\o{}}lmer}}, \ and\
  \bibinfo {author} {\bibfnamefont {T.}~\bibnamefont {Pohl}},\ }\enquote
  {\bibinfo {title} {Photon-photon interactions in {R}ydberg-atom arrays},}\
  \href {\doibase 10.22331/q-2022-03-30-674} {\bibfield  {journal} {\bibinfo
  {journal} {{Quantum}}\ }\textbf {\bibinfo {volume} {6}},\ \bibinfo {pages}
  {674} (\bibinfo {year} {2022})}\BibitemShut {NoStop}%
\bibitem [{\citenamefont {Clemens}\ \emph {et~al.}(2003)\citenamefont
  {Clemens}, \citenamefont {Horvath}, \citenamefont {Sanders},\ and\
  \citenamefont {Carmichael}}]{few_atoms_charmichael}%
  \BibitemOpen
  \bibfield  {author} {\bibinfo {author} {\bibfnamefont {J.~P.}\ \bibnamefont
  {Clemens}}, \bibinfo {author} {\bibfnamefont {L.}~\bibnamefont {Horvath}},
  \bibinfo {author} {\bibfnamefont {B.~C.}\ \bibnamefont {Sanders}}, \ and\
  \bibinfo {author} {\bibfnamefont {H.~J.}\ \bibnamefont {Carmichael}},\
  }\enquote {\bibinfo {title} {Collective spontaneous emission from a line of
  atoms},}\ \href {\doibase 10.1103/PhysRevA.68.023809} {\bibfield  {journal}
  {\bibinfo  {journal} {Phys. Rev. A}\ }\textbf {\bibinfo {volume} {68}},\
  \bibinfo {pages} {023809} (\bibinfo {year} {2003})}\BibitemShut {NoStop}%
\bibitem [{\citenamefont {Carmichael}\ and\ \citenamefont
  {Kim}(2000)}]{few_atoms_charmichael_2}%
  \BibitemOpen
  \bibfield  {author} {\bibinfo {author} {\bibfnamefont {H.}~\bibnamefont
  {Carmichael}}\ and\ \bibinfo {author} {\bibfnamefont {K.}~\bibnamefont
  {Kim}},\ }\enquote {\bibinfo {title} {A quantum trajectory unraveling of the
  superradiance master equation.}}\ \href {\doibase
  https://doi.org/10.1016/S0030-4018(99)00694-X} {\bibfield  {journal}
  {\bibinfo  {journal} {Optics Communications}\ }\textbf {\bibinfo {volume}
  {179}},\ \bibinfo {pages} {417} (\bibinfo {year} {2000})}\BibitemShut
  {NoStop}%
\bibitem [{\citenamefont {Masson}\ \emph {et~al.}(2020)\citenamefont {Masson},
  \citenamefont {Ferrier-Barbut}, \citenamefont {Orozco}, \citenamefont
  {Browaeys},\ and\ \citenamefont {Asenjo-Garcia}}]{Ana_superradiance_1}%
  \BibitemOpen
  \bibfield  {author} {\bibinfo {author} {\bibfnamefont {S.~J.}\ \bibnamefont
  {Masson}}, \bibinfo {author} {\bibfnamefont {I.}~\bibnamefont
  {Ferrier-Barbut}}, \bibinfo {author} {\bibfnamefont {L.~A.}\ \bibnamefont
  {Orozco}}, \bibinfo {author} {\bibfnamefont {A.}~\bibnamefont {Browaeys}}, \
  and\ \bibinfo {author} {\bibfnamefont {A.}~\bibnamefont {Asenjo-Garcia}},\
  }\enquote {\bibinfo {title} {Many-Body Signatures of Collective Decay in
  Atomic Chains},}\ \href {\doibase 10.1103/PhysRevLett.125.263601} {\bibfield
  {journal} {\bibinfo  {journal} {Phys. Rev. Lett.}\ }\textbf {\bibinfo
  {volume} {125}},\ \bibinfo {pages} {263601} (\bibinfo {year}
  {2020})}\BibitemShut {NoStop}%
\bibitem [{\citenamefont {Masson}\ and\ \citenamefont
  {Asenjo-Garcia}(2022)}]{Ana_superradiance_2}%
  \BibitemOpen
  \bibfield  {author} {\bibinfo {author} {\bibfnamefont {S.~J.}\ \bibnamefont
  {Masson}}\ and\ \bibinfo {author} {\bibfnamefont {A.}~\bibnamefont
  {Asenjo-Garcia}},\ }\enquote {\bibinfo {title} {Universality of Dicke
  superradiance in arrays of quantum emitters},}\ \href {\doibase
  10.1038/s41467-022-29805-4} {\bibfield  {journal} {\bibinfo  {journal}
  {Nature Communications}\ }\textbf {\bibinfo {volume} {13}},\ \bibinfo {pages}
  {2285} (\bibinfo {year} {2022})}\BibitemShut {NoStop}%
\bibitem [{\citenamefont {Sierra}\ \emph {et~al.}(2022)\citenamefont {Sierra},
  \citenamefont {Masson},\ and\ \citenamefont
  {Asenjo-Garcia}}]{Ana_superradiance_3}%
  \BibitemOpen
  \bibfield  {author} {\bibinfo {author} {\bibfnamefont {E.}~\bibnamefont
  {Sierra}}, \bibinfo {author} {\bibfnamefont {S.~J.}\ \bibnamefont {Masson}},
  \ and\ \bibinfo {author} {\bibfnamefont {A.}~\bibnamefont {Asenjo-Garcia}},\
  }\enquote {\bibinfo {title} {Dicke Superradiance in Ordered Lattices:
  Dimensionality Matters},}\ \href {\doibase 10.1103/PhysRevResearch.4.023207}
  {\bibfield  {journal} {\bibinfo  {journal} {Phys. Rev. Research}\ }\textbf
  {\bibinfo {volume} {4}},\ \bibinfo {pages} {023207} (\bibinfo {year}
  {2022})}\BibitemShut {NoStop}%
\bibitem [{\citenamefont {Robicheaux}(2021)}]{Robicheaux_superradiance}%
  \BibitemOpen
  \bibfield  {author} {\bibinfo {author} {\bibfnamefont {F.}~\bibnamefont
  {Robicheaux}},\ }\enquote {\bibinfo {title} {Theoretical study of early-time
  superradiance for atom clouds and arrays},}\ \href {\doibase
  10.1103/PhysRevA.104.063706} {\bibfield  {journal} {\bibinfo  {journal}
  {Phys. Rev. A}\ }\textbf {\bibinfo {volume} {104}},\ \bibinfo {pages}
  {063706} (\bibinfo {year} {2021})}\BibitemShut {NoStop}%
\bibitem [{\citenamefont {Lin}\ and\ \citenamefont
  {Yelin}(2012)}]{AMO_GuindarLin}%
  \BibitemOpen
  \bibfield  {author} {\bibinfo {author} {\bibfnamefont {G.-D.}\ \bibnamefont
  {Lin}}\ and\ \bibinfo {author} {\bibfnamefont {S.~F.}\ \bibnamefont
  {Yelin}},\ }in\ \href {\doibase
  https://doi.org/10.1016/B978-0-12-396482-3.00006-5} {\emph {\bibinfo
  {booktitle} {Advances in Atomic, Molecular, and Optical Physics}}},\ \bibinfo
  {series} {Advances In Atomic, Molecular, and Optical Physics}, Vol.~\bibinfo
  {volume} {61},\ \bibinfo {editor} {edited by\ \bibinfo {editor}
  {\bibfnamefont {P.}~\bibnamefont {Berman}}, \bibinfo {editor} {\bibfnamefont
  {E.}~\bibnamefont {Arimondo}}, \ and\ \bibinfo {editor} {\bibfnamefont
  {C.}~\bibnamefont {Lin}}}\ (\bibinfo  {publisher} {Academic Press},\ \bibinfo
  {year} {2012})\ pp.\ \bibinfo {pages} {295--329}\BibitemShut {NoStop}%
\bibitem [{\citenamefont {Rubies-Bigorda}\ and\ \citenamefont
  {Yelin}(2021)}]{Oriol_superradiance}%
  \BibitemOpen
  \bibfield  {author} {\bibinfo {author} {\bibfnamefont {O.}~\bibnamefont
  {Rubies-Bigorda}}\ and\ \bibinfo {author} {\bibfnamefont {S.~F.}\
  \bibnamefont {Yelin}},\ }\enquote {\bibinfo {title} {Superradiance and
  subradiance in inverted atomic arrays},}\ \href
  {https://arxiv.org/abs/2110.11288} {\bibfield  {journal} {\bibinfo  {journal}
  {arXiv:2110.11288}\ } (\bibinfo {year} {2021})}\BibitemShut {NoStop}%
\bibitem [{\citenamefont {Ma}\ \emph {et~al.}(2022)\citenamefont {Ma},
  \citenamefont {Rubies-Bigorda},\ and\ \citenamefont {Yelin}}]{Hanzhen}%
  \BibitemOpen
  \bibfield  {author} {\bibinfo {author} {\bibfnamefont {H.}~\bibnamefont
  {Ma}}, \bibinfo {author} {\bibfnamefont {O.}~\bibnamefont {Rubies-Bigorda}},
  \ and\ \bibinfo {author} {\bibfnamefont {S.~F.}\ \bibnamefont {Yelin}},\
  }\enquote {\bibinfo {title} {Superradiance and subradiance in a gas of
  two-level atoms},}\ \href {https://arxiv.org/abs/2205.15255} {\bibfield
  {journal} {\bibinfo  {journal} {arXiv:2205.15255}\ } (\bibinfo {year}
  {2022})}\BibitemShut {NoStop}%
\bibitem [{\citenamefont {Kubo}(1962)}]{Cumulant_Kubo}%
  \BibitemOpen
  \bibfield  {author} {\bibinfo {author} {\bibfnamefont {R.}~\bibnamefont
  {Kubo}},\ }\enquote {\bibinfo {title} {Generalized Cumulant Expansion
  Method},}\ \href {\doibase 10.1143/JPSJ.17.1100} {\bibfield  {journal}
  {\bibinfo  {journal} {Journal of the Physical Society of Japan}\ }\textbf
  {\bibinfo {volume} {17}},\ \bibinfo {pages} {1100} (\bibinfo {year}
  {1962})}\BibitemShut {NoStop}%
\bibitem [{\citenamefont {Kr{\"a}mer}\ and\ \citenamefont
  {Ritsch}(2015)}]{Ritsch_cumulants_dipole}%
  \BibitemOpen
  \bibfield  {author} {\bibinfo {author} {\bibfnamefont {S.}~\bibnamefont
  {Kr{\"a}mer}}\ and\ \bibinfo {author} {\bibfnamefont {H.}~\bibnamefont
  {Ritsch}},\ }\enquote {\bibinfo {title} {Generalized mean-field approach to
  simulate the dynamics of large open spin ensembles with long range
  interactions},}\ \href {\doibase 10.1140/epjd/e2015-60266-5} {\bibfield
  {journal} {\bibinfo  {journal} {The European Physical Journal D}\ }\textbf
  {\bibinfo {volume} {69}},\ \bibinfo {pages} {282} (\bibinfo {year}
  {2015})}\BibitemShut {NoStop}%
\bibitem [{\citenamefont {Plankensteiner}\ \emph {et~al.}(2022)\citenamefont
  {Plankensteiner}, \citenamefont {Hotter},\ and\ \citenamefont
  {Ritsch}}]{Ritsch_cumulants_package}%
  \BibitemOpen
  \bibfield  {author} {\bibinfo {author} {\bibfnamefont {D.}~\bibnamefont
  {Plankensteiner}}, \bibinfo {author} {\bibfnamefont {C.}~\bibnamefont
  {Hotter}}, \ and\ \bibinfo {author} {\bibfnamefont {H.}~\bibnamefont
  {Ritsch}},\ }\enquote {\bibinfo {title} {Quantum{C}umulants.jl: {A} {J}ulia
  framework for generalized mean-field equations in open quantum systems},}\
  \href {\doibase 10.22331/q-2022-01-04-617} {\bibfield  {journal} {\bibinfo
  {journal} {{Quantum}}\ }\textbf {\bibinfo {volume} {6}},\ \bibinfo {pages}
  {617} (\bibinfo {year} {2022})}\BibitemShut {NoStop}%
\bibitem [{\citenamefont {Robicheaux}\ and\ \citenamefont
  {Suresh}(2021)}]{Robicheaux_cumulants}%
  \BibitemOpen
  \bibfield  {author} {\bibinfo {author} {\bibfnamefont {F.}~\bibnamefont
  {Robicheaux}}\ and\ \bibinfo {author} {\bibfnamefont {D.~A.}\ \bibnamefont
  {Suresh}},\ }\enquote {\bibinfo {title} {Beyond lowest order mean-field
  theory for light interacting with atom arrays},}\ \href {\doibase
  10.1103/PhysRevA.104.023702} {\bibfield  {journal} {\bibinfo  {journal}
  {Phys. Rev. A}\ }\textbf {\bibinfo {volume} {104}},\ \bibinfo {pages}
  {023702} (\bibinfo {year} {2021})}\BibitemShut {NoStop}%
\bibitem [{\citenamefont {Dum}\ \emph {et~al.}(1992)\citenamefont {Dum},
  \citenamefont {Zoller},\ and\ \citenamefont {Ritsch}}]{dum_monte_1992}%
  \BibitemOpen
  \bibfield  {author} {\bibinfo {author} {\bibfnamefont {R.}~\bibnamefont
  {Dum}}, \bibinfo {author} {\bibfnamefont {P.}~\bibnamefont {Zoller}}, \ and\
  \bibinfo {author} {\bibfnamefont {H.}~\bibnamefont {Ritsch}},\ }\enquote
  {\bibinfo {title} {Monte {Carlo} simulation of the atomic master equation for
  spontaneous emission},}\ \href {\doibase 10.1103/PhysRevA.45.4879} {\bibfield
   {journal} {\bibinfo  {journal} {Physical Review A}\ }\textbf {\bibinfo
  {volume} {45}},\ \bibinfo {pages} {4879} (\bibinfo {year}
  {1992})}\BibitemShut {NoStop}%
\bibitem [{\citenamefont {Mølmer}\ \emph {et~al.}(1993)\citenamefont
  {Mølmer}, \citenamefont {Castin},\ and\ \citenamefont
  {Dalibard}}]{molmer_monte_1993}%
  \BibitemOpen
  \bibfield  {author} {\bibinfo {author} {\bibfnamefont {K.}~\bibnamefont
  {Mølmer}}, \bibinfo {author} {\bibfnamefont {Y.}~\bibnamefont {Castin}}, \
  and\ \bibinfo {author} {\bibfnamefont {J.}~\bibnamefont {Dalibard}},\
  }\enquote {\bibinfo {title} {Monte {Carlo} wave-function method in quantum
  optics},}\ \href {\doibase 10.1364/JOSAB.10.000524} {\bibfield  {journal}
  {\bibinfo  {journal} {Journal of the Optical Society of America B}\ }\textbf
  {\bibinfo {volume} {10}},\ \bibinfo {pages} {524} (\bibinfo {year}
  {1993})}\BibitemShut {NoStop}%
\bibitem [{\citenamefont {Lehmberg}(1970{\natexlab{a}})}]{Lehmberg_1970_1}%
  \BibitemOpen
  \bibfield  {author} {\bibinfo {author} {\bibfnamefont {R.~H.}\ \bibnamefont
  {Lehmberg}},\ }\enquote {\bibinfo {title} {Radiation from an $N$-Atom System.
  I. General Formalism},}\ \href {\doibase 10.1103/PhysRevA.2.883} {\bibfield
  {journal} {\bibinfo  {journal} {Phys. Rev. A}\ }\textbf {\bibinfo {volume}
  {2}},\ \bibinfo {pages} {883} (\bibinfo {year}
  {1970}{\natexlab{a}})}\BibitemShut {NoStop}%
\bibitem [{\citenamefont {Lehmberg}(1970{\natexlab{b}})}]{Lehmberg_1970_2}%
  \BibitemOpen
  \bibfield  {author} {\bibinfo {author} {\bibfnamefont {R.~H.}\ \bibnamefont
  {Lehmberg}},\ }\enquote {\bibinfo {title} {Radiation from an $N$-Atom System.
  II. Spontaneous Emission from a Pair of Atoms},}\ \href {\doibase
  10.1103/PhysRevA.2.889} {\bibfield  {journal} {\bibinfo  {journal} {Phys.
  Rev. A}\ }\textbf {\bibinfo {volume} {2}},\ \bibinfo {pages} {889} (\bibinfo
  {year} {1970}{\natexlab{b}})}\BibitemShut {NoStop}%
\bibitem [{\citenamefont {Rubies-Bigorda}\ \emph
  {et~al.}(2022{\natexlab{b}})\citenamefont {Rubies-Bigorda}, \citenamefont
  {Ostermann},\ and\ \citenamefont {Yelin}}]{our_PRL_subradiance}%
  \BibitemOpen
  \bibfield  {author} {\bibinfo {author} {\bibfnamefont {O.}~\bibnamefont
  {Rubies-Bigorda}}, \bibinfo {author} {\bibfnamefont {S.}~\bibnamefont
  {Ostermann}}, \ and\ \bibinfo {author} {\bibfnamefont {S.~F.}\ \bibnamefont
  {Yelin}},\ }\enquote {\bibinfo {title} {Generating multi-excitation
  subradiant states in incoherently excited atomic arrays},}\ \href
  {https://arxiv.org/abs/2209.00034} {\bibfield  {journal} {\bibinfo  {journal}
  {arXiv:2209.00034}\ } (\bibinfo {year} {2022}{\natexlab{b}})}\BibitemShut
  {NoStop}%
\bibitem [{\citenamefont {Huang}\ \emph {et~al.}(2022)\citenamefont {Huang},
  \citenamefont {Yuan}, \citenamefont {Holman}, \citenamefont {Kwon},
  \citenamefont {Masson}, \citenamefont {Gutierrez-Jauregui}, \citenamefont
  {Asenjo-Garcia}, \citenamefont {Will},\ and\ \citenamefont
  {Yu}}]{tweezers_Columbia}%
  \BibitemOpen
  \bibfield  {author} {\bibinfo {author} {\bibfnamefont {X.}~\bibnamefont
  {Huang}}, \bibinfo {author} {\bibfnamefont {W.}~\bibnamefont {Yuan}},
  \bibinfo {author} {\bibfnamefont {A.}~\bibnamefont {Holman}}, \bibinfo
  {author} {\bibfnamefont {M.}~\bibnamefont {Kwon}}, \bibinfo {author}
  {\bibfnamefont {S.~J.}\ \bibnamefont {Masson}}, \bibinfo {author}
  {\bibfnamefont {R.}~\bibnamefont {Gutierrez-Jauregui}}, \bibinfo {author}
  {\bibfnamefont {A.}~\bibnamefont {Asenjo-Garcia}}, \bibinfo {author}
  {\bibfnamefont {S.}~\bibnamefont {Will}}, \ and\ \bibinfo {author}
  {\bibfnamefont {N.}~\bibnamefont {Yu}},\ }\enquote {\bibinfo {title}
  {Metasurface Holographic Optical Traps for Ultracold Atoms},}\ \href
  {https://arxiv.org/abs/2210.07425} {\bibfield  {journal} {\bibinfo  {journal}
  {arXiv preprint arXiv:2210.07425}\ } (\bibinfo {year} {2022})}\BibitemShut
  {NoStop}%
\bibitem [{\citenamefont {Andersen}\ \emph {et~al.}(2021)\citenamefont
  {Andersen}, \citenamefont {Gelly}, \citenamefont {Scuri}, \citenamefont
  {Dwyer}, \citenamefont {Wild}, \citenamefont {Bekenstein}, \citenamefont
  {Sushko}, \citenamefont {Sung}, \citenamefont {Zhou}, \citenamefont {Zibrov}
  \emph {et~al.}}]{trond_andersen}%
  \BibitemOpen
  \bibfield  {author} {\bibinfo {author} {\bibfnamefont {T.~I.}\ \bibnamefont
  {Andersen}}, \bibinfo {author} {\bibfnamefont {R.~J.}\ \bibnamefont {Gelly}},
  \bibinfo {author} {\bibfnamefont {G.}~\bibnamefont {Scuri}}, \bibinfo
  {author} {\bibfnamefont {B.~L.}\ \bibnamefont {Dwyer}}, \bibinfo {author}
  {\bibfnamefont {D.~S.}\ \bibnamefont {Wild}}, \bibinfo {author}
  {\bibfnamefont {R.}~\bibnamefont {Bekenstein}}, \bibinfo {author}
  {\bibfnamefont {A.}~\bibnamefont {Sushko}}, \bibinfo {author} {\bibfnamefont
  {J.}~\bibnamefont {Sung}}, \bibinfo {author} {\bibfnamefont {Y.}~\bibnamefont
  {Zhou}}, \bibinfo {author} {\bibfnamefont {A.~A.}\ \bibnamefont {Zibrov}},
  \emph {et~al.},\ }\enquote {\bibinfo {title} {Spatial light modulation at the
  nanosecond scale with an atomically thin reflector},}\ \href
  {https://arxiv.org/abs/2111.04781} {\bibfield  {journal} {\bibinfo  {journal}
  {arXiv:2111.04781}\ } (\bibinfo {year} {2021})}\BibitemShut {NoStop}%
\bibitem [{\citenamefont {Dimer}\ \emph {et~al.}(2007)\citenamefont {Dimer},
  \citenamefont {Estienne}, \citenamefont {Parkins},\ and\ \citenamefont
  {Carmichael}}]{dicke_cavity_proposal_Carmichael}%
  \BibitemOpen
  \bibfield  {author} {\bibinfo {author} {\bibfnamefont {F.}~\bibnamefont
  {Dimer}}, \bibinfo {author} {\bibfnamefont {B.}~\bibnamefont {Estienne}},
  \bibinfo {author} {\bibfnamefont {A.~S.}\ \bibnamefont {Parkins}}, \ and\
  \bibinfo {author} {\bibfnamefont {H.~J.}\ \bibnamefont {Carmichael}},\
  }\enquote {\bibinfo {title} {Proposed realization of the Dicke-model quantum
  phase transition in an optical cavity QED system},}\ \href {\doibase
  10.1103/PhysRevA.75.013804} {\bibfield  {journal} {\bibinfo  {journal} {Phys.
  Rev. A}\ }\textbf {\bibinfo {volume} {75}},\ \bibinfo {pages} {013804}
  (\bibinfo {year} {2007})}\BibitemShut {NoStop}%
\bibitem [{\citenamefont {Baumann}\ \emph {et~al.}(2010)\citenamefont
  {Baumann}, \citenamefont {Guerlin}, \citenamefont {Brennecke},\ and\
  \citenamefont {Esslinger}}]{baumann2010dicke}%
  \BibitemOpen
  \bibfield  {author} {\bibinfo {author} {\bibfnamefont {K.}~\bibnamefont
  {Baumann}}, \bibinfo {author} {\bibfnamefont {C.}~\bibnamefont {Guerlin}},
  \bibinfo {author} {\bibfnamefont {F.}~\bibnamefont {Brennecke}}, \ and\
  \bibinfo {author} {\bibfnamefont {T.}~\bibnamefont {Esslinger}},\ }\enquote
  {\bibinfo {title} {Dicke quantum phase transition with a superfluid gas in an
  optical cavity},}\ \href {\doibase 10.1038/nature09009} {\bibfield  {journal}
  {\bibinfo  {journal} {Nature}\ }\textbf {\bibinfo {volume} {464}},\ \bibinfo
  {pages} {1301} (\bibinfo {year} {2010})}\BibitemShut {NoStop}%
\bibitem [{\citenamefont {Klinder}\ \emph {et~al.}(2015)\citenamefont
  {Klinder}, \citenamefont {Keßler}, \citenamefont {Wolke}, \citenamefont
  {Mathey},\ and\ \citenamefont
  {Hemmerich}}]{dynamical_phase_transitions_dicke_Hemmerich}%
  \BibitemOpen
  \bibfield  {author} {\bibinfo {author} {\bibfnamefont {J.}~\bibnamefont
  {Klinder}}, \bibinfo {author} {\bibfnamefont {H.}~\bibnamefont {Keßler}},
  \bibinfo {author} {\bibfnamefont {M.}~\bibnamefont {Wolke}}, \bibinfo
  {author} {\bibfnamefont {L.}~\bibnamefont {Mathey}}, \ and\ \bibinfo {author}
  {\bibfnamefont {A.}~\bibnamefont {Hemmerich}},\ }\enquote {\bibinfo {title}
  {Dynamical phase transition in the open Dicke model},}\ \href {\doibase
  10.1073/pnas.1417132112} {\bibfield  {journal} {\bibinfo  {journal}
  {Proceedings of the National Academy of Sciences}\ }\textbf {\bibinfo
  {volume} {112}},\ \bibinfo {pages} {3290} (\bibinfo {year}
  {2015})}\BibitemShut {NoStop}%
\bibitem [{\citenamefont {Ferioli}\ \emph {et~al.}(2022)\citenamefont
  {Ferioli}, \citenamefont {Glicenstein}, \citenamefont {Ferrier-Barbut},\ and\
  \citenamefont {Browaeys}}]{ferioli2022observation}%
  \BibitemOpen
  \bibfield  {author} {\bibinfo {author} {\bibfnamefont {G.}~\bibnamefont
  {Ferioli}}, \bibinfo {author} {\bibfnamefont {A.}~\bibnamefont
  {Glicenstein}}, \bibinfo {author} {\bibfnamefont {I.}~\bibnamefont
  {Ferrier-Barbut}}, \ and\ \bibinfo {author} {\bibfnamefont {A.}~\bibnamefont
  {Browaeys}},\ }\enquote {\bibinfo {title} {Observation of a non-equilibrium
  superradiant phase transition in free space},}\ \href
  {https://arxiv.org/abs/2207.10361} {\bibfield  {journal} {\bibinfo  {journal}
  {arXiv:2207.10361}\ } (\bibinfo {year} {2022})}\BibitemShut {NoStop}%
\bibitem [{\citenamefont {Chew}(1995)}]{GreensFunction_Chew}%
  \BibitemOpen
  \bibfield  {author} {\bibinfo {author} {\bibfnamefont {W.~C.}\ \bibnamefont
  {Chew}},\ }\href@noop {} {\emph {\bibinfo {title} {Waves and Fields in
  Inhomogeneous Media (Electromagnetic Waves)}}}\ (\bibinfo  {publisher} {IEEE
  Computer Society Press},\ \bibinfo {year} {1995})\BibitemShut {NoStop}%
\bibitem [{\citenamefont {Novotny}\ and\ \citenamefont
  {Hecht}(2006)}]{GreensFunction_novotny_hecht_2006}%
  \BibitemOpen
  \bibfield  {author} {\bibinfo {author} {\bibfnamefont {L.}~\bibnamefont
  {Novotny}}\ and\ \bibinfo {author} {\bibfnamefont {B.}~\bibnamefont
  {Hecht}},\ }\href {\doibase 10.1017/CBO9780511813535} {\emph {\bibinfo
  {title} {Principles of Nano-Optics}}}\ (\bibinfo  {publisher} {Cambridge
  University Press},\ \bibinfo {year} {2006})\BibitemShut {NoStop}%
\end{thebibliography}%
\end{document}